\documentclass[preprint,12pt]{elsarticle}

\usepackage{amssymb}

\journal{Annals of Physics}

\begin{document}

\begin{frontmatter}
 
\title{How the antisymmetrization affects a cluster-cluster interaction: two-cluster systems
}


\author[BITP]{V. S. Vasilevsky}
\author[BITP]{Yu. A. Lashko\corref{correspondingauthor}}
\cortext[correspondingauthor]{Yu. A. Lashko}
\ead{ylashko@gmail.com}

\address[BITP]{Bogolyubov Institute for Theoretical Physics,\\
Metrolohichna str., 14b, Kiev, 03143, Ukraine}

\begin{abstract}
We study effects of the antisymmetrization on the potential energy of
two-cluster systems. The object of the investigation is the lightest nuclei of
p-shell with a dominant alpha-cluster channel. For this aim we construct
matrix elements of two-cluster potential energy between cluster oscillator
functions with and without full antisymmetrization. Eigenvalues and
eigenfunctions of the potential energy matrix are studied in detail.
Eigenfunctions of the potential energy operator are presented in oscillator,
coordinate and momentum spaces. We demonstrate that the Pauli principle
affects more strongly the eigenfunctions than the eigenvalues of the matrix 
and leads to the formation of resonance and trapped states.

\end{abstract}


\end{frontmatter}


\section{Introduction}

The aim of this paper is to investigate effects of the Pauli principle on the
potential energy of a two-cluster system. The object of the investigation is
the lightest nuclei of $p$-shell with a dominant $\alpha$-cluster channel. The
method of the investigation is the algebraic version of the resonating group method
\cite{kn:Fil_Okhr, kn:Fil81}.

The resonating group method (RGM) \cite{1937PhRv...52.1083W,
1937PhRv...52.1107W, kn:Saito77, 1978PhR....47..167T,
1981LNP...145..571T, kn:wilderm_eng} is a universal and efficient
tool for studying properties of two-, three- and many-cluster systems. \ This
method works perfectly for describing both compact many-cluster configurations
and weakly bound states. It is also invaluable for studying different types of
reactions and for predicting cross sections of the reactions which are, for
example, important for the astrophysical applications. The main advantages of
the method are that (i) it takes into account the internal structure of
interacting clusters and (ii) correctly treats the Pauli principle. An
important peculiarity of the RGM is that being applied to two- or
three-cluster systems it reduces in a self-consistent way an $A$-body problem
(where $A$ is the number of nucleons) to a two- or three-body problem. The RGM
strongly relies on the translationally invariant many-body shell model
\cite{1970NuPhA.145..593K}, \cite{Neudachin69E}, as this model supplies the
wave functions describing the internal structure of clusters. Within the
model, the internal antisymmetric wave functions of clusters are constructed
from wave functions of three-dimensional harmonic oscillator. Such wave
functions provide a reliable description of light atomic nuclei (the energy of
the ground and excited states, the root-mean-square charge and mass radii and
so on). It is than naturally to use oscillator wave functions also for
describing the inter-cluster motion. At the beginning (see, for example, Refs.
\cite{RevaiPrepr1975, kn:majl}  and book \cite{kn:moshin}) the
oscillator basis was used to study bound states only by the diagonalization
procedure. When the algebraic version of the resonating group method was
formulated \cite{kn:Fil_Okhr}, \cite{kn:Fil81}, the oscillator basis have been
used equally well for studying continuous spectrum states of many-channel and
many-cluster systems.

The Pauli principle and a nucleon-nucleon interaction play the huge role in
the rigorous realizations of the RGM such as the algebraic version of the RGM,
the microscopic $R$ matrix model \cite{1977NuPhA.291..230B} or the no-core
shell model combined with the resonating group method
\cite{2009PhRvC..79d4606Q, 2013PhRvC..88c4320Q}. 
A nucleon-nucleon potential and the
antisymmetrization determine the effective cluster-cluster interaction and
thus strongly affects the dynamics of many-cluster systems.

It is well-known that correct treatment of the Pauli principle within
microscopic models leads to bulky analytical expressions and time-consuming
numerical calculations. To make calculations more simple, several alternative
models for approximate treatment of the Pauli principle have been suggested.
The most popular method is the orthogonality condition model (OCM) which was
suggested by Saito \cite{Saito69, kn:Saito77}. This method involves the
folding approximation and orthogonality of the obtained wave functions to the
Pauli forbidden states. The second method was suggested in Refs.
\cite{Kukulin:1976vf, 1976TMP....27..549K, 1978AnPhy.111..330K}
and was called the pseudo-potential method. This method also employs the
folding approximation for interacting clusters, but uses the separable
pseudo-potential to eliminate the Pauli forbidden states.

It seems that everything is known about the Pauli principle. The Pauli
principle has been investigated many times and from different points of view.
A number of recipes have been formulated for constructing antisymmetric wave
functions for different systems of identical fermions. Many results have been
obtained revealing effects of the antisymmetrization on the structure of bound
states and dynamics of reactions in many-particle systems. Huge efforts have
been applied to study effects of the Pauli principle in two- and many-cluster
systems. A number of publications is devoted to calculation of eigenvalues of
the norm kernel and construction of the Pauli allowed states
\cite{1980PThPh..63..895F, 1977PThPh..58..204H,
1981PThPh..65.1632F, 1981PThPh..65.1901F,
1983PThPh..70..809F, 1984PThPh..72.1277H,
1988PThPh..80..663K, 2003FBS....33..173F,
2004PhRvC..70f4001F, 2005EChAYa..36.1373F,
2009NuPhA.826...24L, 2008NuPhA.806..124L}. Meanwhile, we are
going to demonstrate that some interesting properties of the Pauli principle
remained hidden and we are going to reveal some new intriguing features. We
will demonstrate how the antisymmetrization affects a cluster-cluster
potential energy.

As a tool for this study we employ the method suggested in Ref.
\cite{LASHKO2019167930}. As in Ref. \cite{LASHKO2019167930}, we will construct
matrix\ of potential energy between oscillator functions and then analyze the
eigenvalues and eigenfunctions of the matrix. The eigenfunctions will be
analyzed in the oscillator, coordinate and momentum representations. Involving
three different spaces allows us to get more complete picture on the nature
and properties of the potential energy eigenfunctions. Besides, the method
suggested in \cite{LASHKO2019167930} allows one to reduce a nonlocal
inter-cluster interaction to a local or separable form. It is than interesting
to study what type of a local inter-cluster potential is provided by the
diagonalization procedure. We employ three different nucleon-nucleon
potentials which are often used in many-cluster calculations. These potentials
will help us to demonstrate how eigenvalues and eigenfunctions of the
potential energy operator depend on the shape of nucleon-nucleon potential.

It is important to notice that the total cluster-cluster interaction in a
two-cluster system originates from the nucleon-nucleon interaction and also
from the kinetic energy operator. The influence of the Pauli principle on the
kinetic energy of relative motion of two clusters has been investigated in
\ Refs. \cite{2004PhRvC..70f4001F, 2005PPN..36.714F,
2008NuPhA.806..124L, 2009NuPhA.826...24L}. It has been shown that
the kinetic energy of a two-cluster system is not equivalent to the kinetic
energy of a two-nucleon system due to the antisymmetrization effects. The
kinetic energy operator of two-cluster relative motion modified by the Pauli
principle generates an effective interaction between clusters. The main
properties of such interaction strongly correlate with the dependence of the
eigenvalues of the norm kernel on the number of oscillator quanta. If the
eigenvalues of the norm kernel of a two-cluster system approach unity from
above with increasing the number of oscillator quanta, an effective attraction
between cluster arises. At the same time, an effective inter-cluster repulsion
appears in the case of the eigenvalues tending to unity from below. The radius
of this interaction is determined by the range of oscillator quanta where the
eigenvalues of the norm kernel differ from unity, while the intensity of the
interaction depend on the rate of the eigenvalues approach unity. It was shown
in Ref. \cite{2004PhRvC..70f4001F} that the effective interaction generated by
the kinetic energy operator of the relative motion between clusters can be
strong enough to cause a resonance behavior of the phase shift of
cluster-cluster scattering.

In the present paper we consider only the part of the cluster-cluster
potential generated by the nucleon-nucleon potential with the focus on the
Pauli effects.

The layout of the present paper is the following. In Sec. \ref{Sec:Method} we
present a general framework of our investigations. Section \ref{Sec:Results}
is devoted to numerical analysis of eigenvalues and eigenfunctions of
potential energy operator for selected two-cluster systems. Concluding remarks
and outlook are presented in Sec. \ref{Sec:Conclusions}.

\section{Method \label{Sec:Method}}

A wave function of $A$-nucleon systems for the partition $A=A_{1}+A_{2}$ is%
\begin{equation}
\Psi_{LM}=\widehat{\mathcal{A}}\left\{  \left[  \psi_{1}\left(  A_{1}%
,s_{1},b\right)  \psi_{2}\left(  A_{2},s_{2},b\right)  \right]  _{S}%
f_{L}\left(  q\right)  Y_{LM}\left(  \widehat{\mathbf{q}}\right)  \right\}
,\label{eq:101}%
\end{equation}
where $\psi_{\nu}\left(  A_{\nu},s_{\nu},b\right)  $\ is a fully antisymmetric
function, describing internal structure of the $\nu$th cluster, $\widehat
{\mathcal{A}}$\ is the antisymmetrization operator permuting nucleons
belonging to different clusters and $\mathbf{q}$\ is the\ Jacobi vector
determining distance between interacting clusters
\begin{equation}
\mathbf{q}=\sqrt{\frac{A_{1}A_{2}}{A_{1}+A_{2}}}\left[  \frac{1}{A_{1}}%
\sum_{i\in A_{1}}\mathbf{r}_{i}-\frac{1}{A_{2}}\sum_{j\in A_{2}}\mathbf{r}%
_{j}\right]  .\label{eq:102}%
\end{equation}
The wave functions $\psi_{\nu}\left(  A_{\nu},s_{\nu},b\right)  $ depend on
spatial, spin and isospin coordinates of $A_{\nu}$ nucleons. They also depend
on the oscillator length $b$, since they are the lowest functions of the
translation-invariant oscillator shell model.

In \ Eq. (\ref{eq:101}) we assume that we deal with the $s$-clusters only, it
means that $A_{1},A_{2}\leq4$ and the intrinsic orbital momentum of each
cluster equals zero. The total spin $S$ is a vector sum of the individual
spins of clusters $s_{1}$ and $s_{2}$.

If we omit the antisymmetrization operator in (\ref{eq:101}), we have got the
so-called folding approximation
\begin{equation}
\Psi_{LM}^{\left(  F\right)  }=\left[  \psi_{1}\left(  A_{1},s_{1},b\right)
\psi_{2}\left(  A_{2},s_{2},b\right)  \right]  _{S}f_{L}^{\left(  F\right)
}\left(  q\right)  Y_{LM}\left(  \widehat{\mathbf{q}}\right)  . \label{eq:103}%
\end{equation}
This approximate form is valid when the distance between clusters is large and
effects of the Pauli principle are negligibly small. In what follows we will
discuss in detail how large is this distance.

The function $f_{L}^{\left(  F\right)  }\left(  q\right)  $ is a solution of
the two-body Schr\"{o}dinger equation%
\begin{equation}
\left\{  \widehat{T}_{q}+V^{\left(  F\right)  }\left(  q\right)  +E^{\left(
th\right)  }-E\right\}  f_{L}^{\left(  F\right)  }\left(  q\right)  =0,
\label{eq:104}%
\end{equation}
where
\[
\widehat{T}_{q}=-\frac{\hbar^{2}}{2m}\left[  \frac{d^{2}}{dq^{2}}+\frac{2}%
{q}\frac{d}{dq}-\frac{L\left(  L+1\right)  }{q^{2}}\right]
\]
is the kinetic energy operator and $V^{\left(  F\right)  }\left(  q\right)  $
is a folding or direct potential which is defined as%
\begin{eqnarray}
V^{\left(  F\right)  }\left(  q\right)   &  =&\int d\mathbf{\tau}%
_{1}d\mathbf{\tau}_{2}\left\vert \psi_{1}\left(  A_{1}\right)  \right\vert
^{2}\left\vert \psi_{2}\left(  A_{2}\right)  \right\vert ^{2}\sum_{i\in A_{1}%
}\sum_{j\in A_{2}}\widehat{V}\left(  \mathbf{r}_{ij}\right) \nonumber\\
&  =&\int d\mathbf{r}_{1}d\mathbf{r}_{2}\rho_{1}\left(  \mathbf{r}_{1}\right)
\rho_{2}\left(  \mathbf{r}_{2}\right)  \widehat{V}\left(  \mathbf{r}%
_{1}-\mathbf{r}_{2}+\sqrt{\frac{A_{1}+A_{2}}{A_{1}A_{2}}}\mathbf{q}\right)  .
\label{eq:105}%
\end{eqnarray}
Here $\rho_{1}\left(  \mathbf{r}_{1}\right)  $ ($\rho_{2}\left(
\mathbf{r}_{2}\right)  $) is a single particle local density of the first
(second) cluster, and $\mathbf{r}_{1}$ ($\mathbf{r}_{2}$) is a coordinate of a
nucleon with respect to the center of mass of the corresponding cluster. For
$s$-nuclei, which are described by the lowest many-body shell model functions,
the folding potential of the Coulomb and NN forces, having Gaussian coordinate
form, has a simple analytical form. Folding potentials have been very often
used in numerous calculations of elastic and inelastic processes.

The exact two-cluster potential is a nonlocal operator%
\begin{equation}
V_{L}^{\left(  E\right)  }\left(  \widetilde{q},q\right)  =\left\langle
\widehat{\mathcal{P}}_{L}\left(  \widetilde{q}\right)  \left\vert \widehat
{V}\right\vert \widehat{\mathcal{P}}_{L}\left(  q\right)  \right\rangle ,
\label{eq:111}%
\end{equation}
where%
\begin{equation}
\widehat{V}=\sum_{i<j}\widehat{V}\left(  \mathbf{r}_{ij}\right)
\label{eq:112}%
\end{equation}
is the total potential energy of the system and
\begin{equation}
\widehat{\mathcal{P}}_{L}\left(  q\right)  =\widehat{\mathcal{A}}\left\{
\left[  \psi_{1}\left(  A_{1},s_{1},b\right)  \psi_{2}\left(  A_{2}%
,s_{2},b\right)  \right]  _{S}\delta\left(  r-q\right)  Y_{LM}\left(
\widehat{\mathbf{r}}\right)  \right\}  . \label{eq:113}%
\end{equation}
Operator $\widehat{\mathcal{P}}_{L}\left(  q\right)  $ reduces the
$A$-body space to an effective two-body space.

It is well-known that the inter-cluster wave function $f_{L}\left(  q\right)
$ is a solution to the integro-differential equation. This equation can be
transformed to a simpler form and than can be much easily solved, when\ the
function $f_{L}\left(  q\right)  $ is expanded into a complete set of the
oscillator functions%
\begin{equation}
f_{L}\left(  q\right)  =\sum_{n=0}^{\infty}C_{nL}\Phi_{nL}\left(  q,b\right)
, \label{eq:106}%
\end{equation}
where $C_{nL}$ is the expansion coefficient and $\Phi_{nL}\left(  q,b\right)
$ is the radial part of an oscillator wave function, $n$ is the number of
radial quanta. Oscillator length $b$ for the oscillator function is selected
the same as for many-particle oscillator functions describing the internal
structure of interacting clusters.

Note that function $f_{L}\left(  p\right)  $ in momentum space has the
same expansion coefficients%
\begin{equation}
f_{L}\left(  p\right)  =\sum_{n=0}^{\infty}C_{nL}\Phi_{nL}\left(  p,b\right)
.\label{eq:107}%
\end{equation}
Oscillator functions in coordinate and momentum space are very similar%
\begin{equation}
\left\{
\begin{array}
[c]{c}%
\Phi_{nL}\left(  q,b\right)  \\
\Phi_{nL}\left(  p,b\right)
\end{array}
\right\}  =N_{nL}~\rho^{L}e^{-\frac{1}{2}\rho^{2}}L_{n}^{L+1/2}\left(
\rho^{2}\right)  \cdot\left\{
\begin{array}
[c]{c}%
\left(  -1\right)  ^{n}b^{-3/2}\\
b^{3/2}%
\end{array}
\right\}  ,%
\begin{array}
[c]{c}%
\rho=r/b,\\
\rho=pb,
\end{array}
,\label{eq:108}%
\end{equation}
where normalization coefficient $N_{nL}$ is defined as%
\[
N_{nL}=\sqrt{\frac{2~n!}{\Gamma\left(  n+L+3/2\right)  }}.
\]
These functions are orthonormal and obey the completeness relation.

One can find in \cite{kn:cohstate2E},\cite{kn:cohstate1E} all necessary
details of calculating matrix elements of potential energy for nuclei under consideration.

We have to construct the antisymmetric cluster basis functions%
\begin{equation}
\left\vert nL\right\rangle _{C}=\widehat{\mathcal{A}}\left\{  \left[  \psi
_{1}\left(  A_{1},s_{1},b\right)  \psi_{2}\left(  A_{2},s_{2},b\right)
\right]  _{S}\Phi_{nL}\left(  q,b\right)  Y_{LM}\left(  \widehat{\mathbf{q}%
}\right)  \right\}  \label{eq:201}%
\end{equation}
and the folding cluster basis functions%
\begin{equation}
\left\vert nL\right\rangle _{F}=\left[  \psi_{1}\left(  A_{1},s_{1},b\right)
\psi_{2}\left(  A_{2},s_{2},b\right)  \right]  _{S}\Phi_{nL}\left(
q,b\right)  Y_{LM}\left(  \widehat{\mathbf{q}}\right)  . \label{eq:202}%
\end{equation}
Both sets of functions are complete and orthogonal. In what follows we will
omit labels $s_{1}$, $s_{2}$ and $b$. Folding functions $\left\vert
nL\right\rangle _{F}$ are orthonormal, while antisymmetric cluster basis
functions $\left\vert nL\right\rangle _{C}$ are not normalized to unity:%
\[
\left\langle nL|\widetilde{n}L\right\rangle _{C}=\Lambda_{nL}\delta
_{n,\widetilde{n}},
\]
where $\Lambda_{nL}$ are the well-known eigenvalues of the norm kernel.

By using the cluster basis functions $\left\{  \left\vert nL\right\rangle
_{C}\right\}  $, one obtains the two-cluster Schr\"{o}dinger equation in the
form%
\begin{equation}
\sum_{m=0}\left\{  \left\langle nL\left\vert \widehat{H}\right\vert
mL\right\rangle _{C}-E\Lambda_{nL}\delta_{n,m}\right\}  C_{mL}=0,
\label{eq:203}%
\end{equation}
where $\left\langle nL\left\vert \widehat{H}\right\vert mL\right\rangle _{C}$
is a matrix element of a microscopic two-cluster Hamiltonian. If we introduce
an orthonormal set of the antisymmetric cluster functions%
\begin{equation}
\left\vert nL\right\rangle _{E}=\frac{1}{\sqrt{\Lambda_{nL}}}\left\vert
nL\right\rangle _{C}=\frac{1}{\sqrt{\Lambda_{nL}}}\widehat{\mathcal{A}%
}\left\{  \psi_{1}\left(  A_{1}\right)  \psi_{2}\left(  A_{2}\right)
\Phi_{nL}\left(  q,b\right)  Y_{LM}\left(  \widehat{\mathbf{q}}\right)
\right\}  , \label{eq:201N}%
\end{equation}
we will get the Schr\"{o}dinger equation (\ref{eq:203}) in the form
\begin{equation}
\sum_{m=0}^{\infty}\left\{  \left\langle nL\left\vert \widehat{H}\right\vert
mL\right\rangle _{E}-E\delta_{n,m}\right\}  C_{mL}=0, \label{eq:204}%
\end{equation}
which is typical for an orthonormal basis of functions.

In what follows, it is assumed the energy of two-cluster systems is determined
with respect to the two-cluster threshold. Hence the internal kinetic and
potential energy of interacting clusters is subtracted from the Hamiltonian
$\widehat{H}$ of a two-cluster system. \textit{\ }Therefore, both exact
$V_{L}^{\left(  E\right)  }\left(  q,q\right)  $\textit{\ }and folding
$V^{\left(  F\right)  }\left(  q\right)  $\textit{\ }potential energies tend
to zero as the coordinate\textit{ }$q$\textit{\ }approaches infinity and thus
both of them represent the potential energy of cluster-cluster interaction.
Moreover, the exact\textit{ }$V_{L}^{\left(  E\right)  }\left(  \widetilde
{q},q\right)  $ potential energy for large values of \textit{\ }$\widetilde
{q}$ and $q$ becomes local and coincides with the folding potential
$V^{\left(  F\right)  }\left(  q\right)  $.

The first effect of the Pauli principle on two-cluster systems is connected
with appearance of the Pauli forbidden states. The Pauli forbidden states are
those cluster basis functions\ (\ref{eq:201}) which are annihilated by the
antisymmetrization operator. The eigenvalues of the norm kernel corresponding
to the Pauli forbidden states are equal to zero. Thus the Pauli principle
forbids such basis functions \cite{1988PThPh..80..663K},
\cite{1968PThPh..40..893S}, \cite{Saito69}, \cite{kn:Saito77}. The Pauli
forbidden states are the cluster basis functions with the lowest values of the
quantum number $n$. The number of \ the forbidden states depends on the
clusterization and the total orbital momentum $L$ of a compound system. For
example, for the $0^{+}$ state in $^{8}$Be$=\alpha+\alpha$, the Pauli
principle annihilates two cluster functions $\left\vert n=0,L=0\right\rangle
_{C}$ and $\left\vert n=1,L=0\right\rangle _{C}$, and thus two first columns
and rows of the potential energy matrix equal zero. \ 

The second effect of the Pauli principle can be seen in Eq. (\ref{eq:201N}).
It renormalizes the two-cluster oscillator functions and, consequently, matrix
elements of the potential energy operator as
\begin{equation}
\left\langle nL\left\vert \widehat{V}\right\vert mL\right\rangle _{E}=\frac
{1}{\sqrt{\Lambda_{nL}}}\left\langle nL\left\vert \widehat{V}\right\vert
mL\right\rangle _{C}\frac{1}{\sqrt{\Lambda_{mL}}}. \label{eq:206}%
\end{equation}
It is obvious that if $\Lambda_{nL}>1$, the matrix elements $\left\langle
nL\left\vert \widehat{V}\right\vert mL\right\rangle _{E}$ are decreased with
respect to the matrix elements $\left\langle nL\left\vert \widehat
{V}\right\vert mL\right\rangle _{C}$. If $\Lambda_{nL}<1$, the matrix elements
$\left\langle nL\left\vert \widehat{V}\right\vert mL\right\rangle _{E}$ are increased.

We will use the set of functions (\ref{eq:201N}) to calculate matrix elements
of potential energy with the exact treatment of the Pauli principle. It is
interesting to analyze how the eigenvalues of the norm kernel change potential
energy of two-cluster system.

\subsection{Matrix elements}

Having constructed matrix of potential energy $\left\Vert \left\langle
nL\left\vert \widehat{V}\right\vert mL\right\rangle \right\Vert $\ of
dimension $N\times N$, we can calculate eigenvalues $\lambda_{\alpha}$
($\alpha$=1, 2, \ldots, $N$) and corresponding eigenfunctions $\left\{
U_{n}^{\alpha}\right\}  $ of the matrix. \ Actually, we use the decomposition
of the matrix into diagonal matrix $\left\Vert \lambda\right\Vert $ \ and
orthogonal one $\left\Vert U\right\Vert $:%
\begin{equation}
\left\Vert \left\langle nL\left\vert \widehat{V}\right\vert mL\right\rangle
\right\Vert =\left\Vert U^{-1}\right\Vert \left\Vert \lambda\right\Vert
\left\Vert U\right\Vert \label{eq:301}%
\end{equation}
or%
\begin{equation}
\left\langle nL\left\vert \widehat{V}\right\vert mL\right\rangle =\sum
_{\alpha=1}^{N}U_{n}^{\alpha}\lambda_{\alpha}U_{m}^{\alpha}, \label{eq:301A}%
\end{equation}
where
\begin{equation}
\left\Vert \lambda\right\Vert =\left\Vert
\begin{array}
[c]{cccc}%
\lambda_{1} &  &  & \\
& \lambda_{2} &  & \\
&  & \ddots & \\
&  &  & \lambda_{N}%
\end{array}
\right\Vert . \label{eq:302}%
\end{equation}
The orthogonal matrix $\left\Vert U\right\Vert $ generates a new set of 
inter-cluster functions $\phi_{\alpha}$ and two-cluster wave functions
$\Psi_{\alpha}$
\begin{eqnarray}
\phi_{\alpha}\left(  q,b\right)   &  =&\sum_{n}U_{n}^{\alpha}\Phi_{nL}\left(
q,b\right) \label{eq:303A}\\
\Psi_{\alpha}  &  =&\widehat{\mathcal{A}}\left\{  \psi_{1}\left(  A_{1}\right)
\psi_{2}\left(  A_{2}\right)  \phi_{\alpha}\left(  q,b\right)  Y_{LM}\left(
\widehat{\mathbf{q}}\right)  \right\}  . \label{eq:303B}%
\end{eqnarray}

The functions $\phi_{\alpha}\left(  q,b\right)  $ and eigenvalues
$\lambda_{\alpha}$ enable us to construct inter-cluster nonlocal potential
\begin{equation}
\widehat{V}_{N}\left(  q,\widetilde{q}\right)  =\sum_{\alpha=1}^{N}%
\phi_{\alpha}\left(  q,b\right)  \lambda_{\alpha}\phi_{\alpha}\left(
\widetilde{q},b\right)  . \label{eq:304}%
\end{equation}
This is an approximate form of the inter-cluster potential. One can get an
exact form of the potential as a limit of the expression%
\[
\widehat{V}\left(  q,\widetilde{q}\right)  =\lim_{N\rightarrow\infty}%
\widehat{V}_{N}\left(  q,\widetilde{q}\right)  .
\]
Note, that for the folding (direct) potential, we have got%
\[
\widehat{V}\left(  q\right)  \delta\left(  q-\widetilde{q}\right)
=\lim_{N\rightarrow\infty}\widehat{V}_{N}\left(  q,\widetilde{q}\right)  .
\]
In what follows we are going to study properties of the eigenvalues and
eigenfunctions of the potential energy operator in the oscillator
representation $\left\{  U_{n}^{\alpha}\right\}  $, coordinate $\phi_{\alpha
}\left(  q,b\right)  $ and momentum $\phi_{\alpha}\left(  p,b\right)  $
spaces. \ 

Note that the formula (\ref{eq:304}) is one of the numerous methods of the
separable representation in quantum mechanics (see review and books about the
separable representations in \cite{Zubarev_EChAYa76}, \cite{bookBelyaev786E},
\cite{kn:Newton}). For a two-body case, the eigenfunctions $\phi_{\alpha
}\left(  q,b\right)  $ or $\phi_{\alpha}\left(  p,b\right)  $ would
immediately define a wave function and t-matrix, as it was demonstrated in
Ref. \cite{LASHKO2019167930}. However, in two-cluster systems the
antisymmetrization is known to affect the kinetic energy and norm kernel and
thus the kinetic energy and norm kernel participate in creating the effective
cluster-cluster interaction as well.

When $L\geq A-4$, eigenfunctions of exact and folding approximation are very
close to each other. However, for small values of $n$ one can see the trace of
the antisymmetrization operator.

\section{Results \label{Sec:Results}}

We employ three nucleon-nucleon potentials which have been often used in
different realizations of the cluster model. In our calculations we involve
the Volkov N2 (VP) \cite{kn:Volk65}, modified Hasegawa-Nagata (MHNP)
\cite{potMHN1, potMHN2} and Minnesota (MP) \cite{kn:Minn_pot1} potentials.
Coulomb forces are also involved in calculations and treated exactly. For the
sake of simplicity we neglect the spin-orbit forces, thus the total spin $S$
and the total orbital momentum $L$ are good quantum numbers. Oscillator length
$b$ is selected to optimize energy of the lowest decay threshold for each nucleus and for each
NN potential. The optimal values of $b$ are shown in Tab. \ref{Tab:Optimb}.%

\begin{table}[tbp] \centering
\caption{Oscillator length $b$ in fm for different nuclei and different
potentials.}%
\begin{tabular}
[c]{|l|l|l|l|}\hline
Nucleus & VP & MHNP & MP\\\hline
$^{5}$He, $^{5}$Li & 1.38 & 1.32 & 1.28\\
$^{6}$Li & 1.46 & 1.36 & 1.31\\
$^{7}$Li, $^{7}$Be & 1.44 & 1.36 & 1.35\\
$^{8}$Be & 1.38 & 1.32 & 1.28\\\hline
\end{tabular}
\label{Tab:Optimb}%
\end{table}%

\begin{figure}[ptb]
\begin{center}
\includegraphics[
width=\textwidth] 
{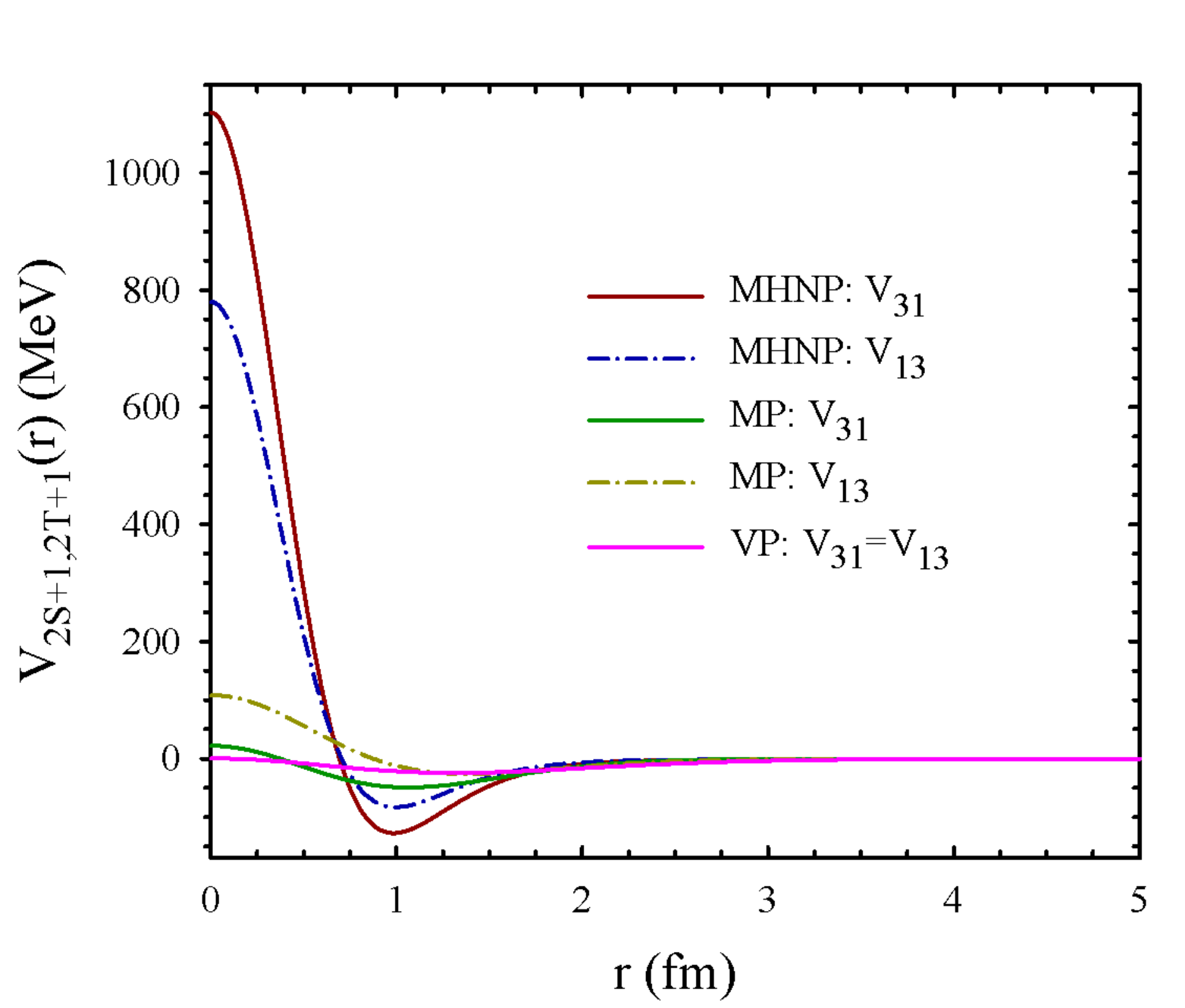}%
\caption{Even components of the MHNP, MP and VP potentials as a function of
distance between nucleons.}%
\label{Fig:NNpotentials}%
\end{center}
\end{figure}

\begin{figure}[ptb]
\begin{center}
\includegraphics[
width=\textwidth] 
{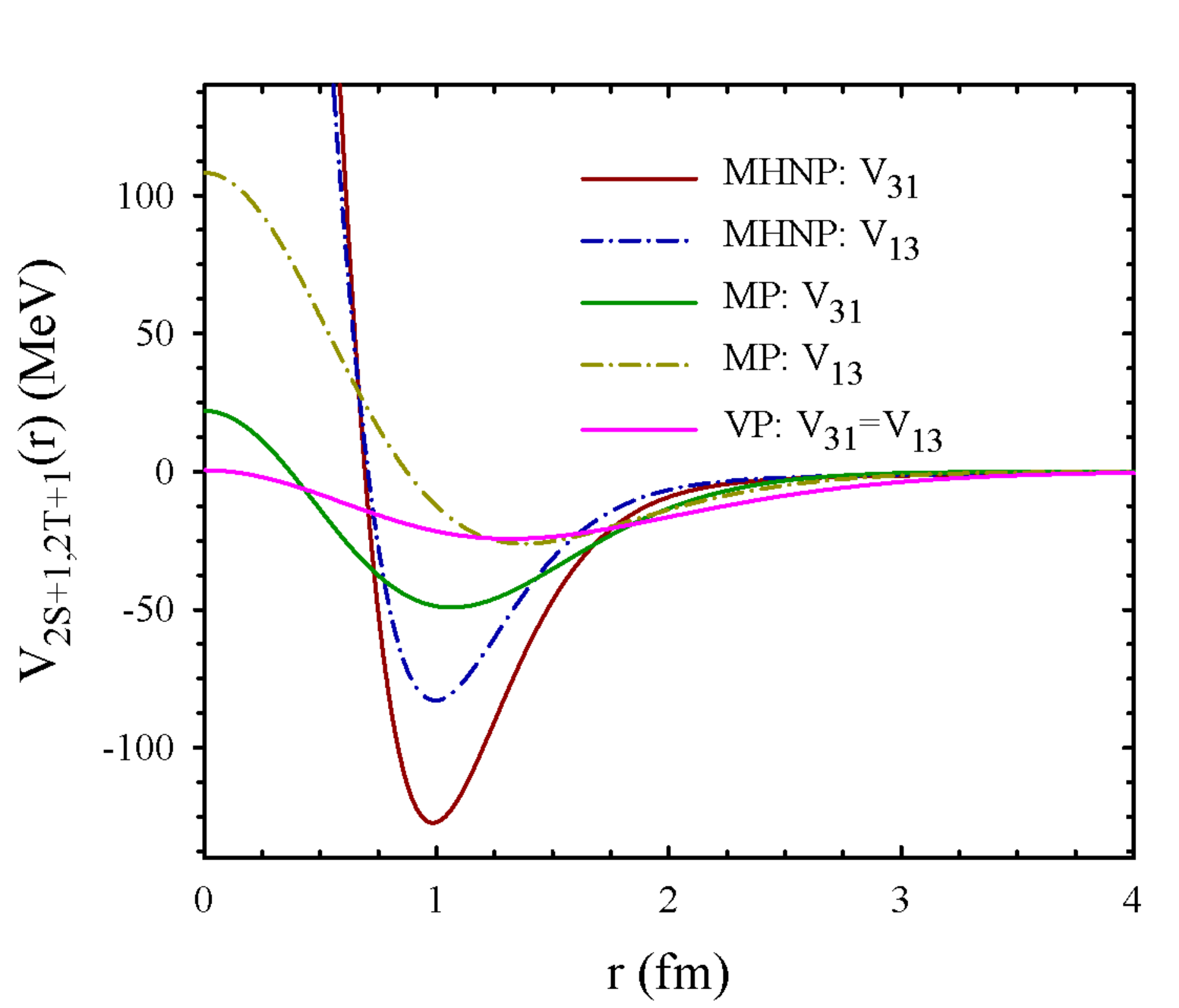}%
\caption{An attractive part of the even components of the MHNP, MP and VP.}%
\label{Fig:NNpotentialsS}%
\end{center}
\end{figure}

Figures \ref{Fig:NNpotentials} and \ref{Fig:NNpotentialsS} demonstrate the
main differences between the MHNP, MP and VP potentials. In these figures we
display only the even components $V_{31}$ and $V_{13}$ of all three NN
potentials. It is known, that the even components are stronger than the odd
components. Indeed, the $V_{31}$ component provides a bound state in a
deuteron and the $V_{13}$ component generates a virtual state in the
neutron-neutron system. One can see that the MHNP has the largest repulsive
core at small distances, the MP has \ a soft core and the VP has a negligibly
small core. Fig. \ref{Fig:NNpotentialsS} provides detailed view of these
potentials. We observe from this figure that the larger is the core in the
potential, the deeper is the attractive part of the potential.%

%

Nuclei under consideration together with their corresponding two-cluster
configurations are listed in Table \ref{Tab:Nucl&Config}. One can see that all
nuclei are represented by the lowest and thus dominant two-cluster channel. \ %

\begin{table}[tbp] \centering
\caption{List of nuclei and two-cluster configurations}%
\begin{tabular}
[c]{|l|l|}\hline
Nucleus & Configuration\\\hline
$^{5}$He & $^{4}$He$+n$\\
$^{5}$Li & $^{4}$He$+p$\\
$^{6}$Li & $^{4}$He$+d$\\
$^{7}$Li & $^{4}$He$+^{3}$H\\
$^{7}$Be & $^{4}$He$+^{3}$He\\
$^{8}$Be & $^{4}$He$+^{4}$He\\\hline
\end{tabular}
\label{Tab:Nucl&Config}%
\end{table}%

In what follows, we will consider matrix elements of potential energy operator
between the Pauli allowed states and will neglect the Pauli forbidden states.
This will lead to\ somewhat different numeration of basis functions
(\ref{eq:201}) and matrix elements (\ref{eq:206}). The quantum number $n$ we
substitute with a new quantum number
\begin{equation}
n\rightarrow n_{0}+n, \label{eq:310}%
\end{equation}
where $n$ \ (n = 0, 1, \ldots) numerates the Pauli allowed states, and
$n_{0}$ is the number of the Pauli forbidden states
\begin{equation}
n_{0}=\left\{
\begin{array}
[c]{cc}%
\max\left(  \left(  A-4-L\right)  /2,0\right)  & \pi=\left(  -\right)
^{A}=\left(  -\right)  ^{L}\\
\max\left(  \left(  A-3-L\right)  /2,0\right)  & \pi=\left(  -\right)
^{A+1}=\left(  -\right)  ^{L}%
\end{array}
\right.  . \label{eq:311}%
\end{equation}
The first row of this equation is valid for normal parity states, while
the second row is valid for abnormal parity states. It is easy to deduce from Eq.
(\ref{eq:311}) that there are no  Pauli forbidden states for the total
angular momentum $L\geq\left(  A-4\right)  $ (the normal parity states) or for
$L\geq\left(  A-3\right)  $ (the abnormal parity states). In the nuclei under
consideration the number of the Pauli forbidden states is very small. For
example, there is only one Pauli forbidden state in the abnormal parity
$0^{+}$ state of $^{5}$He and $^{5}$Li, and there are two Pauli forbidden
states in the $0^{+}$ states of $^{7}$Li, $^{7}$Be and $^{8}$Be.

\subsection{Eigenvalues of the norm kernel.}

\begin{figure}[ptb]
\begin{center}
\includegraphics[
width=\textwidth] 
{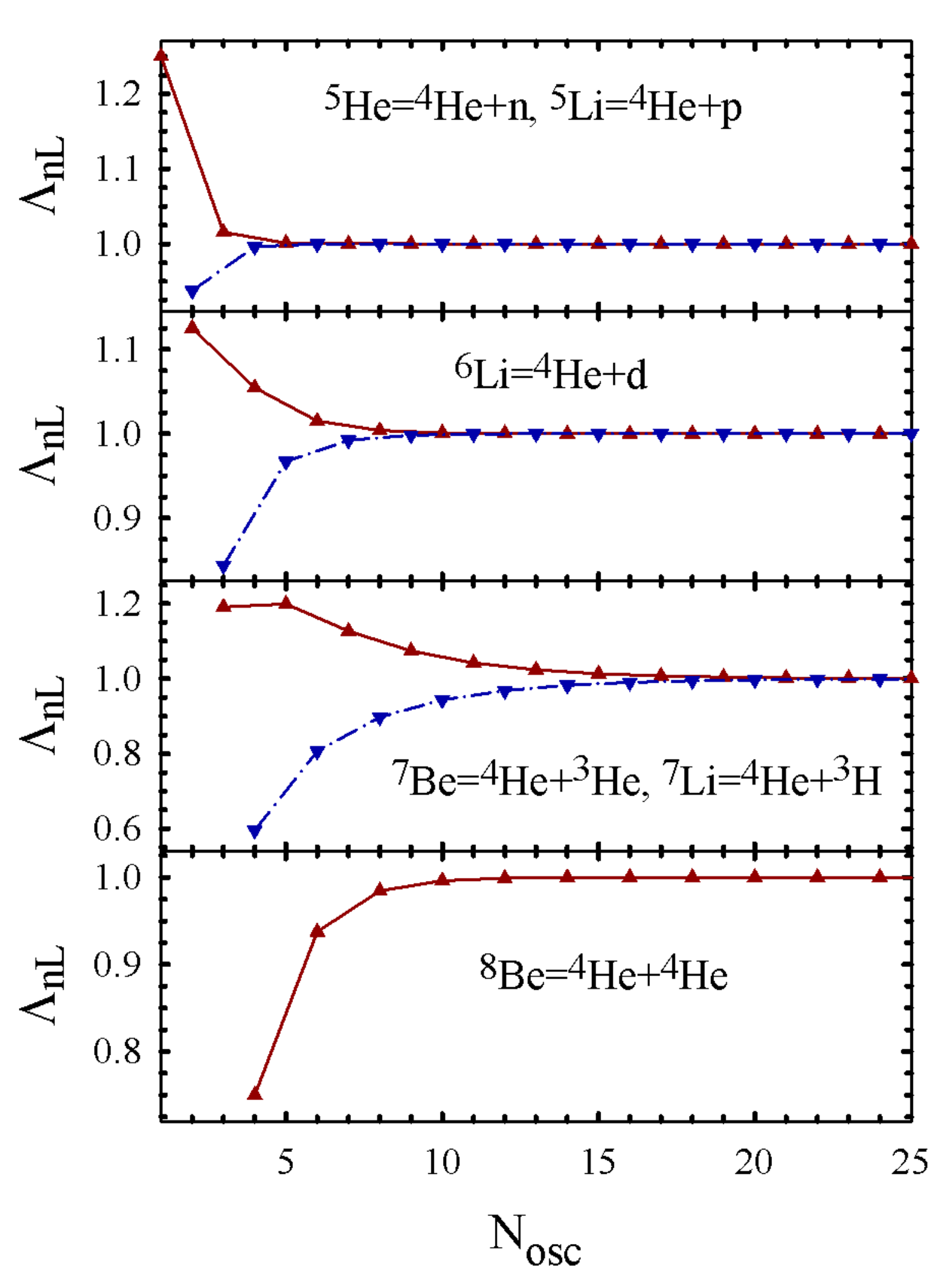}%
\caption{Eigenvalues of the norm kernel for the lightest $p$-shell nuclei.
Triangles up represent the eigenvalues for the normal parity states, and
triangles down demonstrate them for the abnormal parity states.}%
\label{Fig:NormEigenvalN}%
\end{center}
\end{figure}

First, we calculate eigenvalues of the norm kernel. They are displayed in Fig.
\ref{Fig:NormEigenvalN}\ as a function of $N_{osc}$, the number of oscillator
quanta
\[
N_{osc}=2n+L.
\]
The eigenvalues are displayed for normal parity (triangles up) and for
abnormal parity (triangles down) states. One can see that the
antisymmetrization affects only few cluster basis states with $N_{osc}\leq25
$. \ We can estimate the "range" of the Pauli principle by using
correspondence between oscillator and coordinate spaces:
\[
R_{P}\approx b\sqrt{\frac{A_{1}+A_{2}}{A_{1}A_{2}}}\sqrt{2N_{osc}+3}%
\approx10~fm.
\]%

Fig. \ref{Fig:NormEigenvalN} also demonstrates that for nuclei $^{5}$He,
$^{5}$Li, $^{6}$Li, $^{7}$Li, $^{7}$Be the eigenvalues of the norm operator
$\Lambda_{nL}>1$ for the normal parity states and $\Lambda_{nL}<1$ for the
abnormal parity states.

\begin{figure}[ptb]
\begin{center}
\includegraphics[
width=\textwidth] 
{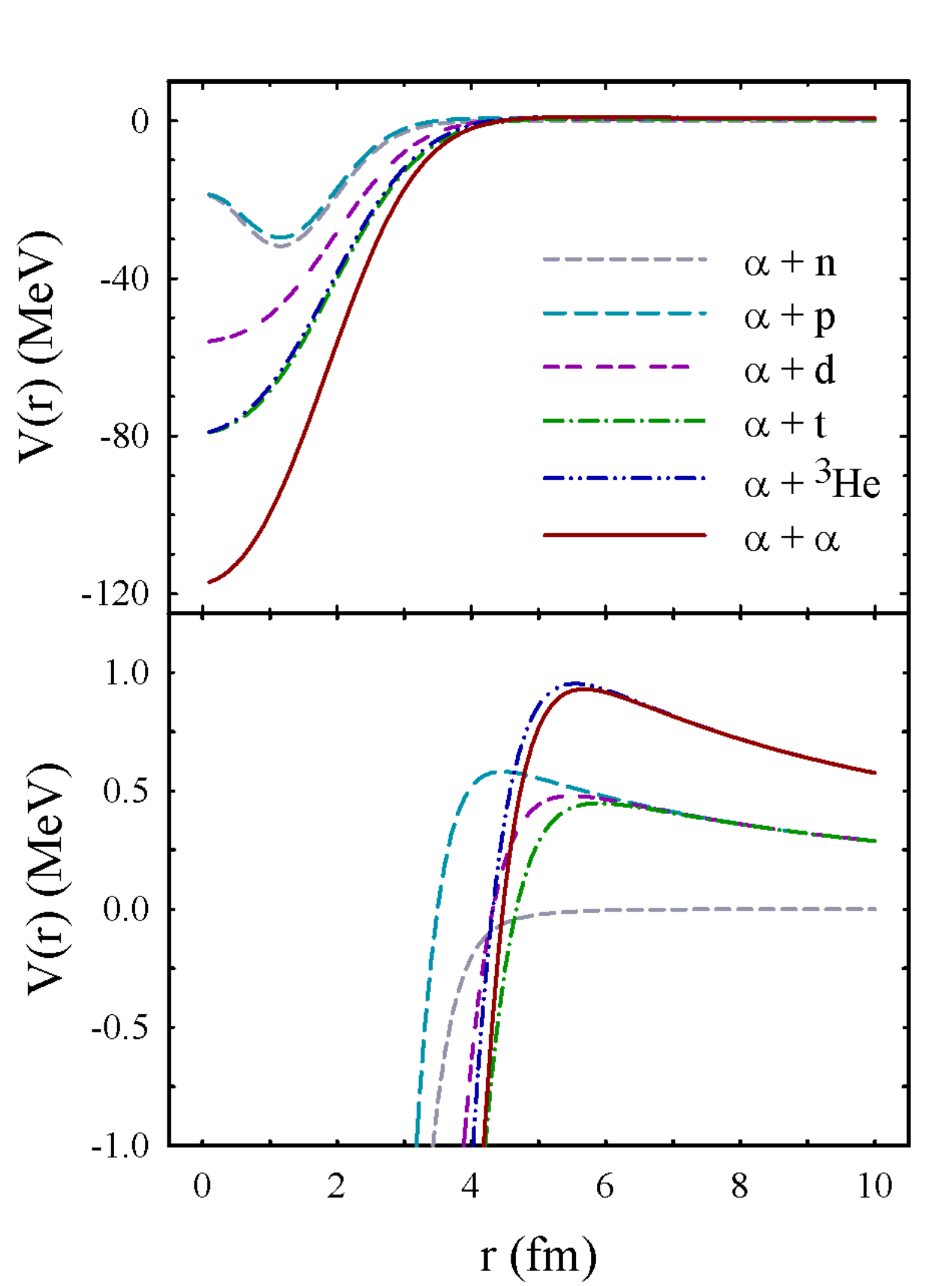}%
\caption{Folding potential generated by the MHN potential as a function of the
\ distance between interacting clusters for the lightest $p$-shell nuclei.}%
\label{Fig:FoldPot_MHN}%
\end{center}
\end{figure}

In Fig \ref{Fig:FoldPot_MHN} we display folding potentials for $^{5}$He,
$^{5}$Li, $^{6}$Li, $^{7}$Li, $^{7}$Be, and $^{8}$Be created by the MHNP
nucleon-nucleon potential and the Coulomb interaction between protons. It is
worthwhile noticing that despite of the huge core in this nucleon-nucleon
potential, there is no such a core in the cluster-cluster folding potential. Only
in $^{5}$He and $^{5}$Li nuclei we can see a small repulsive core at small
inter-cluster distances. The lower part of Fig. \ref{Fig:FoldPot_MHN}
demonstrates the shape and height of a barrier created by the Coulomb
interaction.

\begin{figure}[ptb]
\begin{center}
\includegraphics[
width=\textwidth] 
{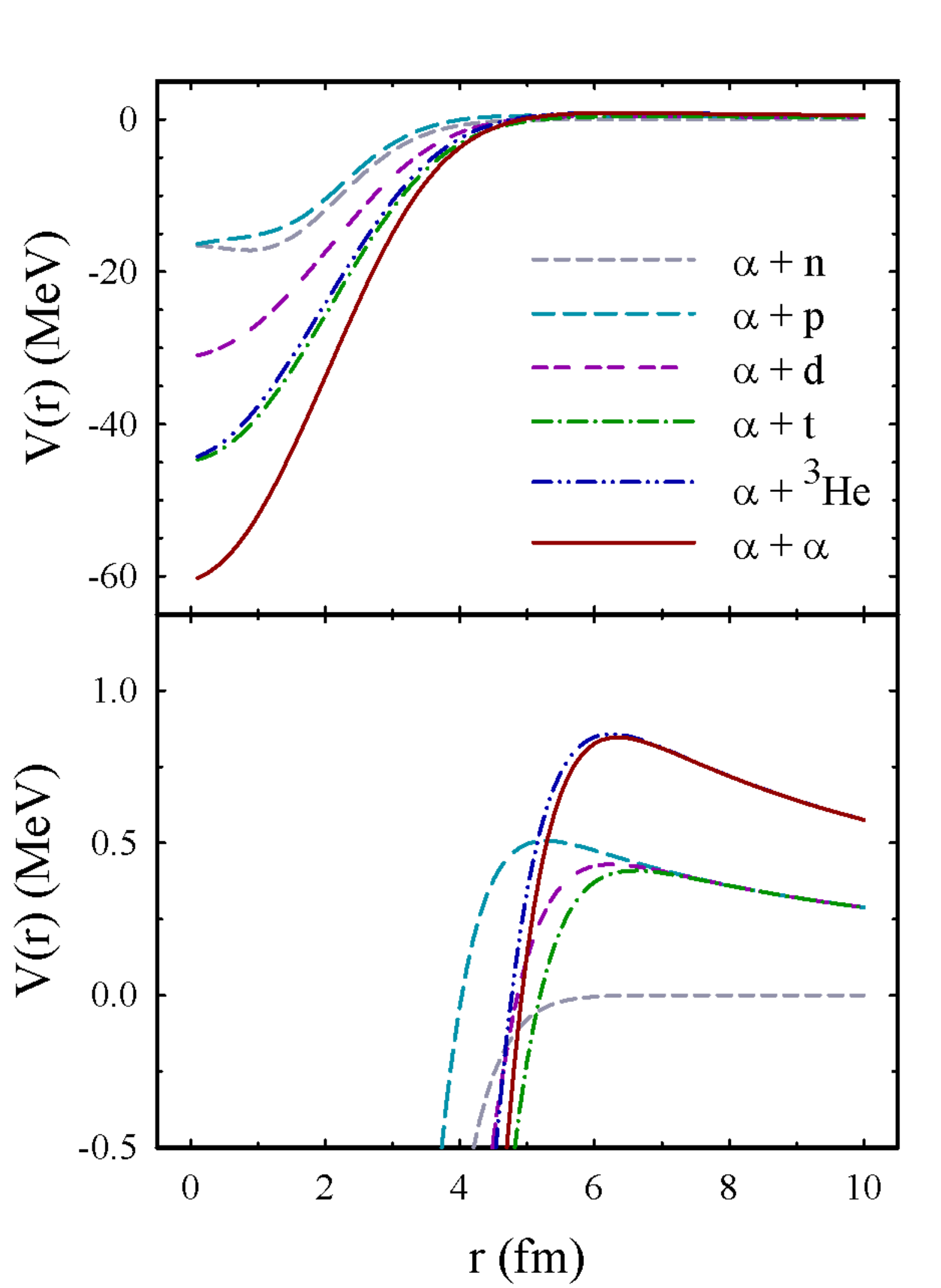}%
\caption{Folding potentials generated by the VP.}%
\label{Fig:FoldPot_VP}%
\end{center}
\end{figure}

In Fig. \ref{Fig:FoldPot_VP}\ we display the folding potentials created by the
VP. As we pointed out above, this nucleon-nucleon potential has no repulsive
core. However, it creates a small repulsive core between $\alpha$-particle and a neutron in $^{5}$He. Different NN
potentials create approximately the same shape of the cluster-cluster folding potential. The
main difference between the folding potentials, generated by different NN
potentials, is depths and width of the potential well.%

\subsection{Matrix of potential energy.}

\begin{figure}[ptb]
\begin{center}
\includegraphics[
width=\textwidth] 
{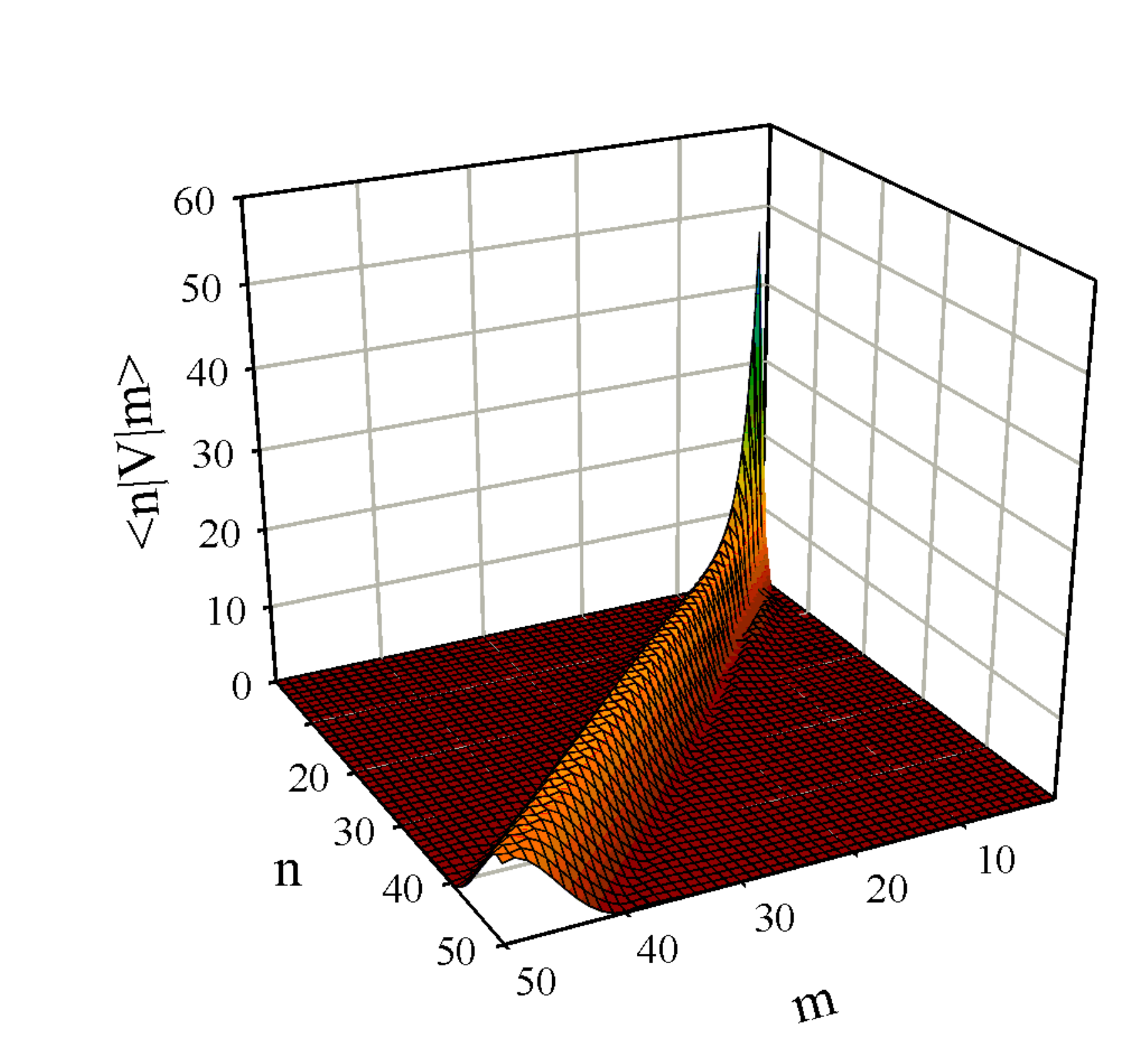}%
\caption{Matrix elements of the exact potential energy operator calculated for
the $0^{+}$ state of $^{8}$Be with the MHNP.}%
\label{Fig:PotEnL0MHNP3D}%
\end{center}
\end{figure}

\begin{figure}[ptb]
\begin{center}
\includegraphics[
width=\textwidth] 
{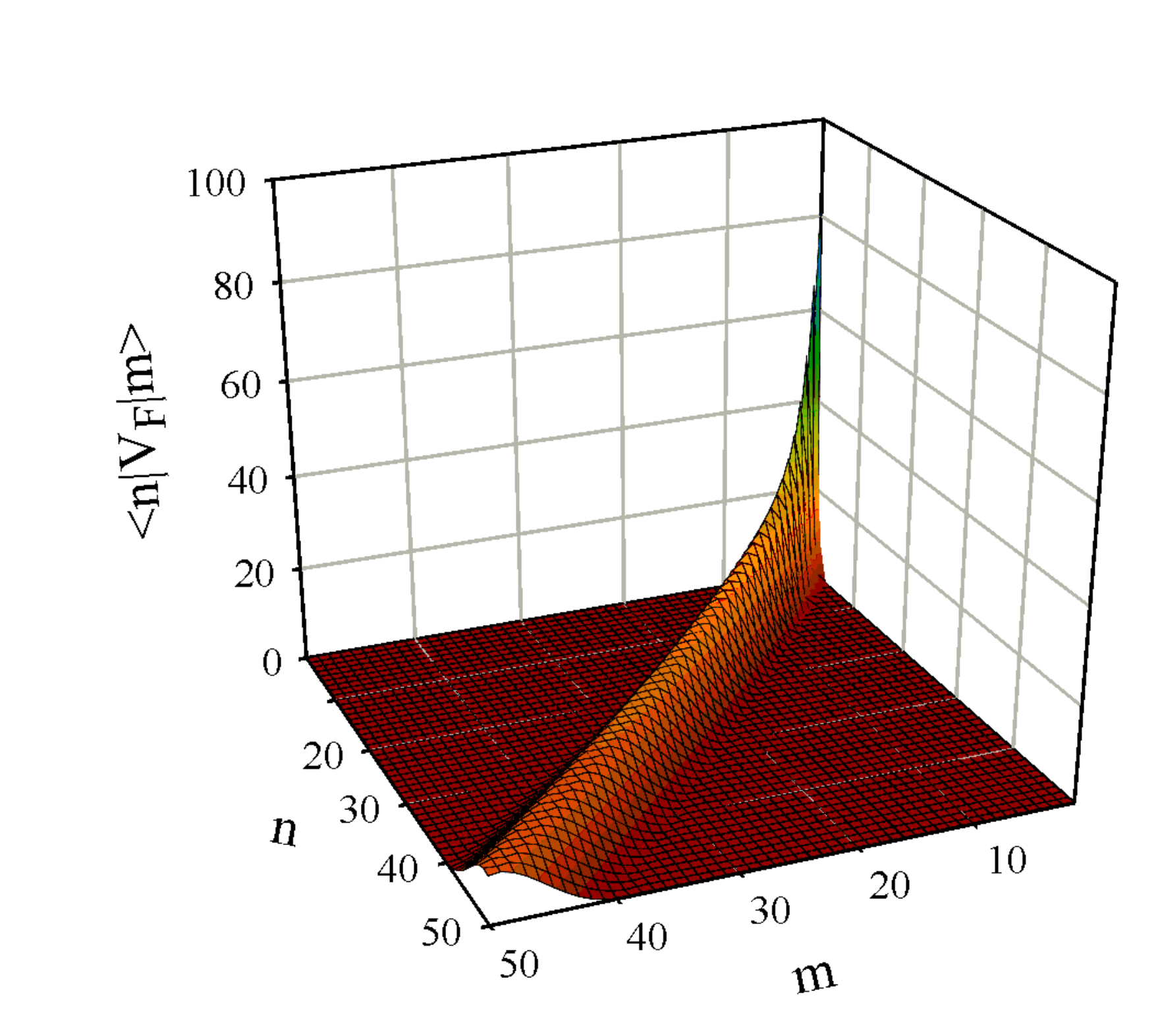}%
\caption{Matrix elements of the folding potential energy operator calculated
for the $0^{+}$ state of $^{8}$Be with the MHNP.}%
\label{Fig:PotEFnL0MHNP3D}%
\end{center}
\end{figure}

In Figs. \ref{Fig:PotEnL0MHNP3D} and \ref{Fig:PotEFnL0MHNP3D} we demonstrate
general features of matrix of the exact and folding potentials, respectively.
These matrices are calculated for the $0^{+}$ state in $^{8}$Be with the MHNP
potential and look very similar, except the region of very small number of
quanta. Matrix elements of the exact and folding potentials have the same
structure. As evident from Figs. \ref{Fig:PotEnL0MHNP3D} and
\ref{Fig:PotEFnL0MHNP3D}, nonzero matrix elements are concentrated around the
main diagonal of the potential matrix. Along this diagonal the matrix elements are
slowly decreasing. This is a general behavior of matrix elements of the
potential energy operator for all nuclei under considerations, for all values
of the total orbital momentum $L$ and for all nucleon-nucleon potentials
involved in calculations.%

%

\subsection{Eigenvalues}%

\begin{figure}[ptb]
\begin{center}
\includegraphics[
width=\textwidth] 
{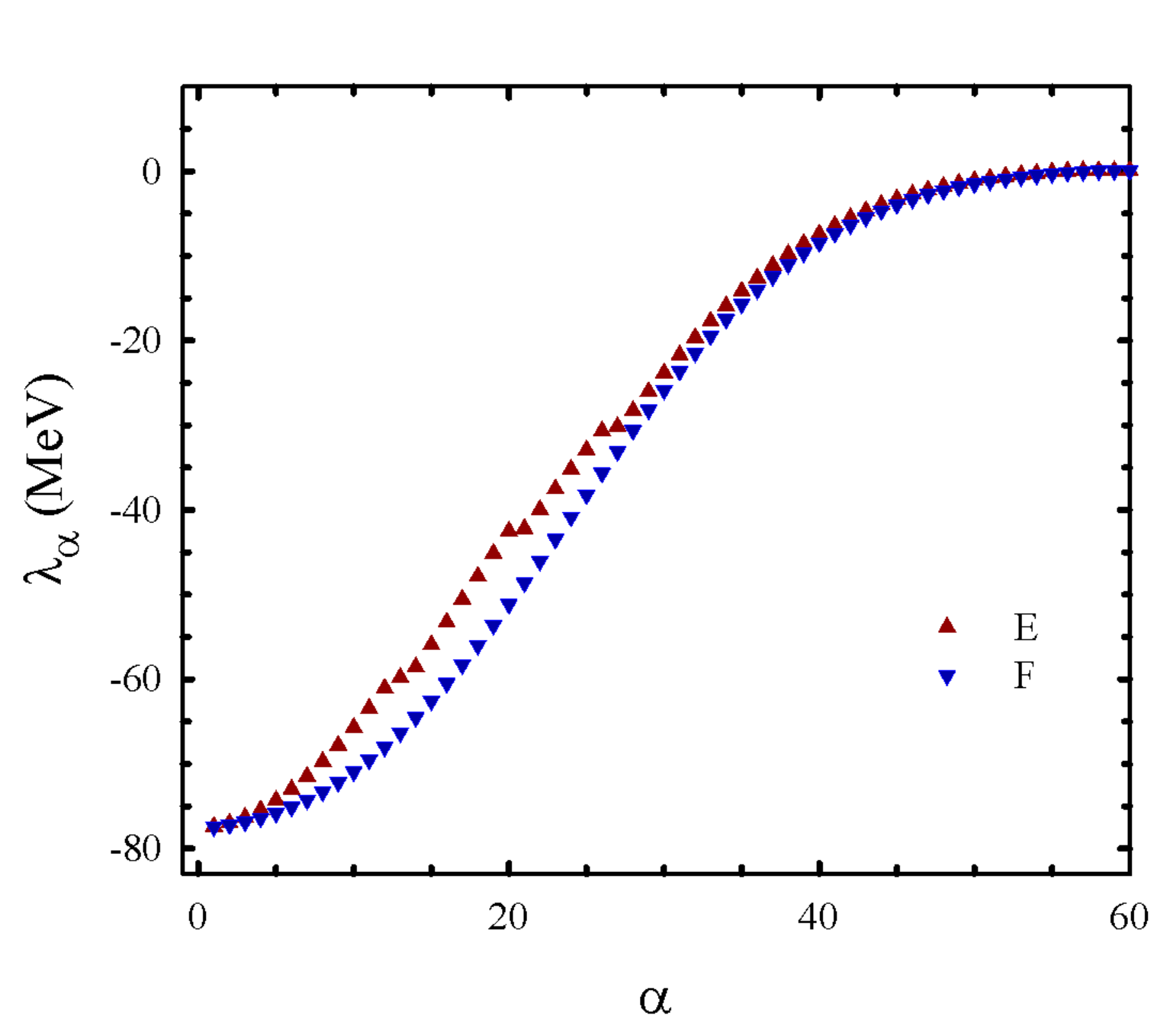}%
\caption{Eigenvalues of the exact (E) and folding (F) potential energy matrix
for the $1^{-}$ state of $^{7}$Li. Results are obtained with the MHNP.}%
\label{Fig:Li7_la_vs_a_L_1_MHN}%
\end{center}
\end{figure}
In Fig \ref{Fig:Li7_la_vs_a_L_1_MHN} we display eigenvalues of the potential
energy matrix for $L^{\pi}=1^{-}$ state of $^{7}$Li, calculated for the MHNP
potential with 300 basis functions. One can see that the eigenvalues of the
potential energy matrix, calculated with antisymmetrization (in what follows
we will mark it with the letter $E$), are very close to the ones determined
in the folding approximation (we mark them with the letter $F$). The lowest
eigenvalues almost coincide, indicating that both potentials have the same
depth.  One can also see that the exact potential is less attractive in the
range $5\leq\alpha\leq30$. Besides, the eigenvalues of the exact potential
reveal some irregularities, which are absent for the folding potential
eigenvalues. Below it will be shown that they correspond to resonance states
generated by the Pauli exclusion principle. For $\alpha>50$ the exact
potential is very close to the folding potential. Similar behavior of
\ eigenvalues is observed for all lightest nuclei of the $p$-shell and for all
NN potentials involved in our calculations.

\begin{figure}[ptbh]
\begin{center}
\includegraphics[
width=\textwidth] 
{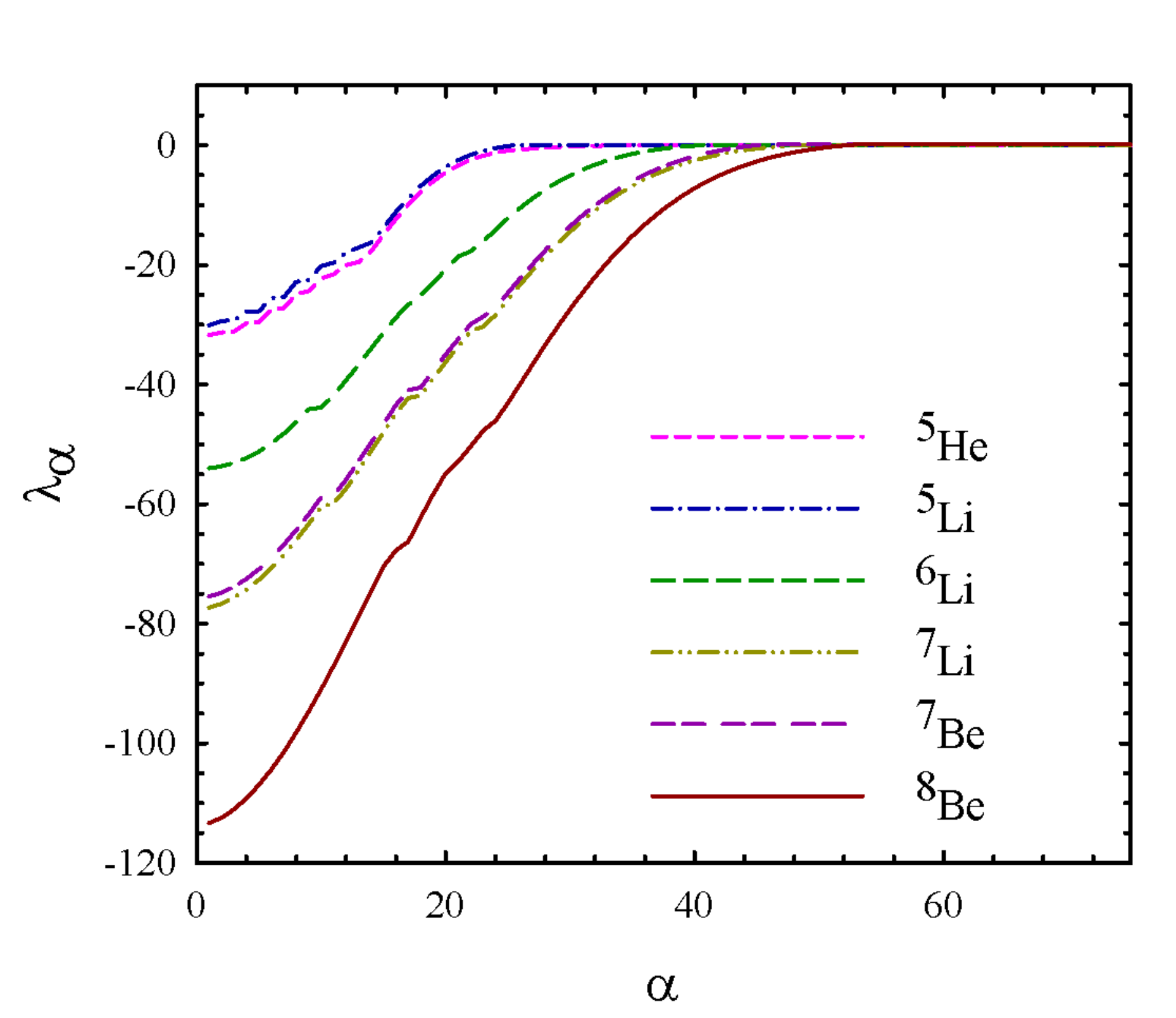}%
\caption{Eigenvalues of the potential energy matrix for the ground state of
$p$-shell nuclei. Results are obtained with the MHNP.}%
\label{Fig:Psh_la_vs_a_MHN}%
\end{center}
\end{figure}

In Fig. \ref{Fig:Psh_la_vs_a_MHN} we show eigenvalues $\lambda_{\alpha}$ for
all lightest $p$-shell nuclei obtained with the MHNP. The eigenvalues are
displayed for the "ground states" of these nuclei. This means that the total
orbital momentum $L=0$ for even nuclei ($^{6}$Li and $^{8}$Be) and $L=1$ for
odd nuclei ($^{5}$He, $^{5}$Li, $^{7}$Li and $^{7}$Be).

\begin{figure}[ptbh]
\begin{center}
\includegraphics[
width=\textwidth] 
{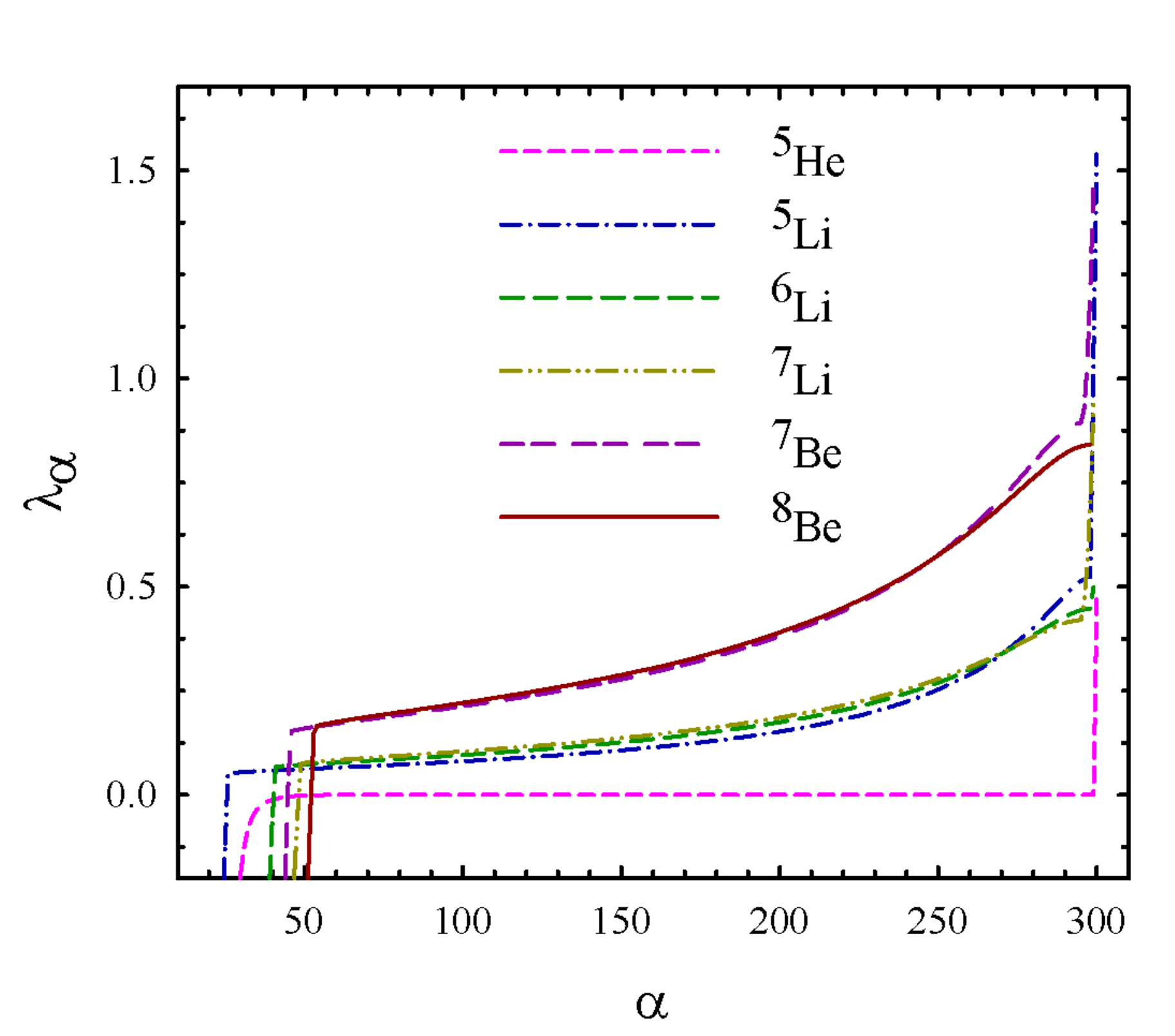}%
\caption{The eigenvalues $\lambda_{\alpha}$ for large values of $\alpha$
generated by the MHNP.}%
\label{Fig:Psh_la_vs_a_MHN_L}%
\end{center}
\end{figure}

One can see that the heavier is a nucleus, the deeper is the effective
potential. One also notices, by comparing results for $^{5}$He and $^{5}$Li,
$^{7}$Li and $^{7}$Be, that the Coulomb interaction slightly modifies the
eigenvalues $\lambda_{\alpha}$ for moderate values of $\alpha<50$. However,
the strongest effect we observe for maximal values of $\alpha$. Indeed, in
Fig. \ref{Fig:Psh_la_vs_a_MHN_L} we display $\lambda_{\alpha}$ which are
positive. We assume that these eigenvalues for nuclei $^{6}$Li, $^{7}$Li,
$^{7}$Be and $^{8}$Be are originated from the Coulomb interaction. It will be
proved later by comparing the eigenvalues of the exact and folding potential
energy operators. Here, we point out that the eigenvalues of the potential
energy of nuclei having the same total charge $Z$ and cluster charges $Z_{1}$
and $Z_{2}$ are very close to each other. Indeed, the lines representing
nuclei $^{5}$Li, $^{6}$Li and $^{7}$Li lie very close. The same is true for
the nuclei $^{7}$Be and $^{8}$Be.

As we can see in Fig. \ref{Fig:Psh_la_vs_a_MHN_L}, for large values of
$\alpha$, the exact eigenvalues $\lambda_{\alpha}$ are very close to the
folding eigenvalues. However, in the vicinity of the final value of
$\alpha=300$, we observe a noticeable difference of eigenvalues $\lambda
_{\alpha}$. Thus we need a closer look at the structure of these unusual
eigenstates, for which the abbreviation HES (highly-excited eigenstate) is used.

\begin{figure}[ptb]
\begin{center}
\includegraphics[
width=\textwidth] 
{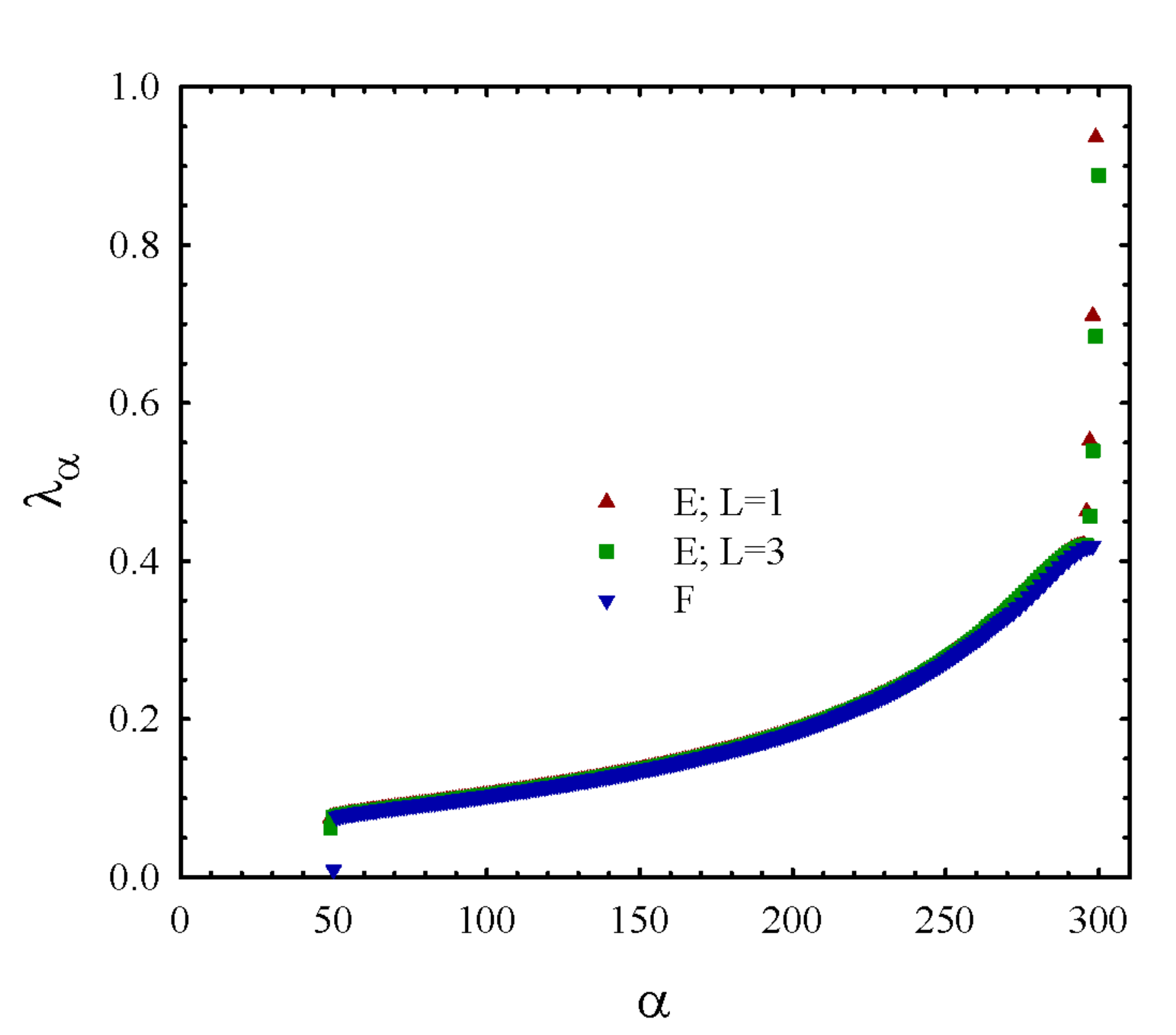}%
\caption{The largest eigenvalues generated by the MHNP for the $1^{-}$ and
$3^{-}$ states in $^{7}$Li. }%
\label{Fig:LEignVal7LiL13MHNP}%
\end{center}
\end{figure}

Having analyzed eigenvalues of all nuclei, we came to the conclusion that  
HES are observed for the normal parity states of all nuclei but $^{8}$Be. For
the abnormal parity states we found out the HES only in $^{5}$He and $^{5}$Li.
The number of the HES depends on the nucleus and is independent on the
nucleon-nucleon potential. For $^{5}$He, $^{5}$Li, $^{6}$Li, there is only one
HES, and for $^{7}$Li and $^{7}$Be the number of such states equals 4. In
Fig. \ref{Fig:LEignVal7LiL13MHNP} we demonstrate the eigenvalues
$\lambda_{\alpha}$ with $\alpha\geq50$. This figure shows the
eigenvalues $\lambda_{\alpha}$ for the $1^{-}$ and $3^{-}$ states in $^{7}$Li,
obtained with and without antisymmetrization. In the neighborhood of a point
$\alpha$=300 the four HES appear in the $1^{-}$ and $3^{-}$ states. The
eigenvalues $\lambda_{\alpha}$ of these state slightly depend on the total
orbital momentum $L$.%

Detailed analysis of wave functions of the HES will be presented at the end of
this section.

\begin{figure}
[ptb]
\begin{center}
\includegraphics[
width=\textwidth] 
{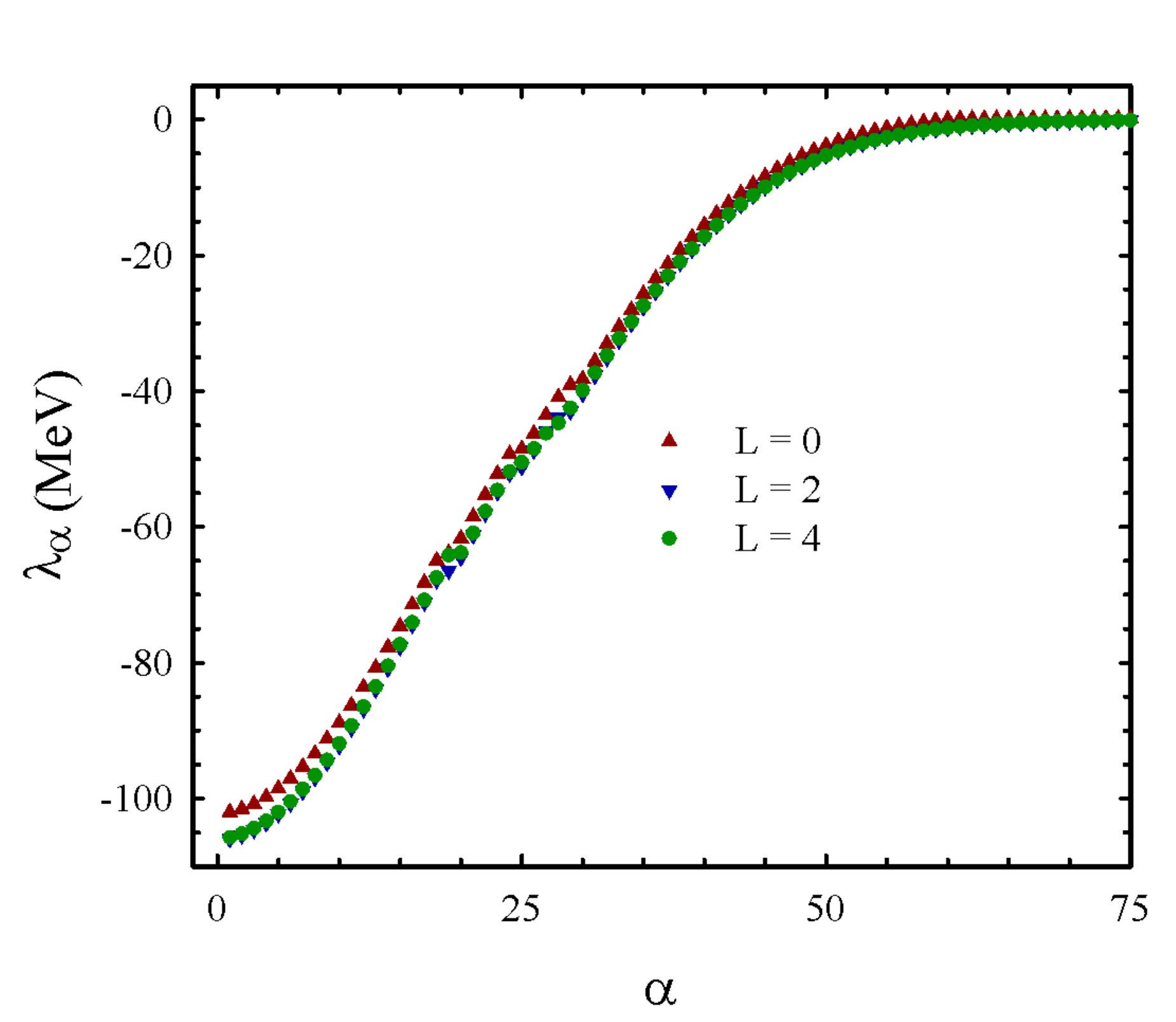}%
\caption{The eigenvalues of matrixes of potential energy operator generated by
the MHNP potential for $^{8}$Be with the total angular momentum $L$=0, 2 and
4.}%
\label{Fig:EigVal8BeL04}%
\end{center}
\end{figure}

Fig. \ref{Fig:EigVal8BeL04} demonstrates how the eigenvalues of the potential
energy operator depend on the total orbital momentum. In Fig.
\ref{Fig:EigVal8BeL04} we display the eigenvalues for $^{8}$Be obtained with
the MHNP. We see that the eigenvalues of the exact potential energy operator
slightly depend on the total orbital momentum. A similar picture is observed
for other nuclei and for other nucleon-nucleon potentials.%

\begin{figure}[ptb]
\begin{center}
\includegraphics[
width=\textwidth] 
{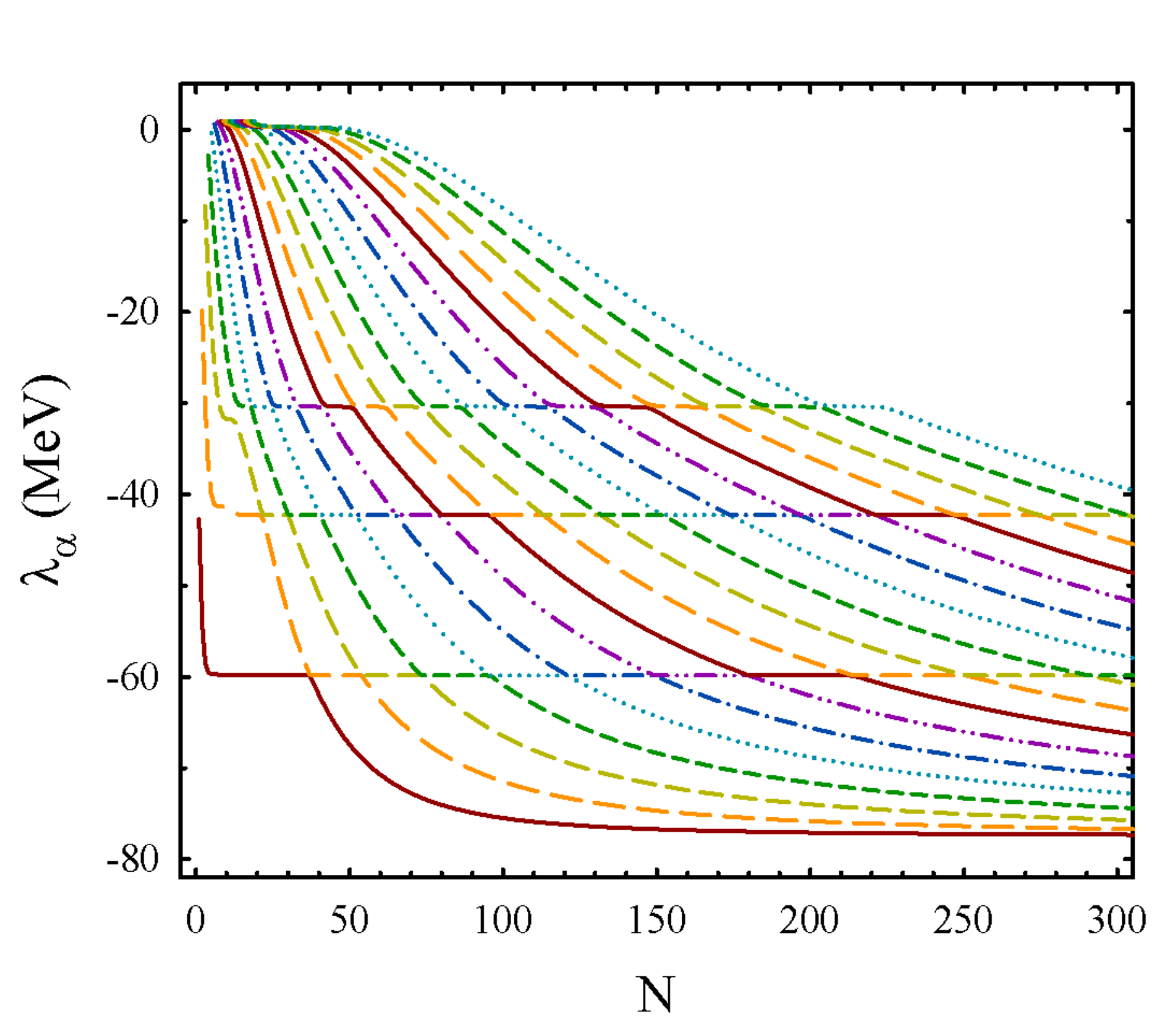}%
\caption{Eigenvalues of the potential energy matrix \ as a function of the
number $N$ of oscillator functions involved in calculations. Results are
obtained for the $1^{-}$ state in $^{7}$Li with the MHNP.}%
\label{Fig:Li7_la_vs_a_L_1_MHNN}%
\end{center}
\end{figure}

In Fig. \ref{Fig:Li7_la_vs_a_L_1_MHNN} we show dependence of the eigenvalues
$\lambda_{\alpha}$ on the number $N$ of oscillator functions involved in
calculations. These results are obtained for $L^{\pi}=1^{-}$ state of \ $^{7}%
$Li with the MHNP.%

One can see that the more functions involved, the more dense is the spectrum
of eigenstates. One also notices three plateaus created by the eigenstates
which can be treated as "resonance" states. It is because such plateau in the
dependence of the eigenvalues of  a Hamiltonian indicate that a system under
consideration has resonance states. They are called the Harris eigenstates
\cite{1967PhRvL..19..173H, 1984JMP....25..317Y}. The energy of such
plateau is used to locate the position of a narrow resonance state in the
stabilization method \cite{1970PhRvA...1.1109H}. We will discuss properties of
such states in the next subsection.

\begin{figure}[ptb]
\begin{center}
\includegraphics[
width=\textwidth] 
{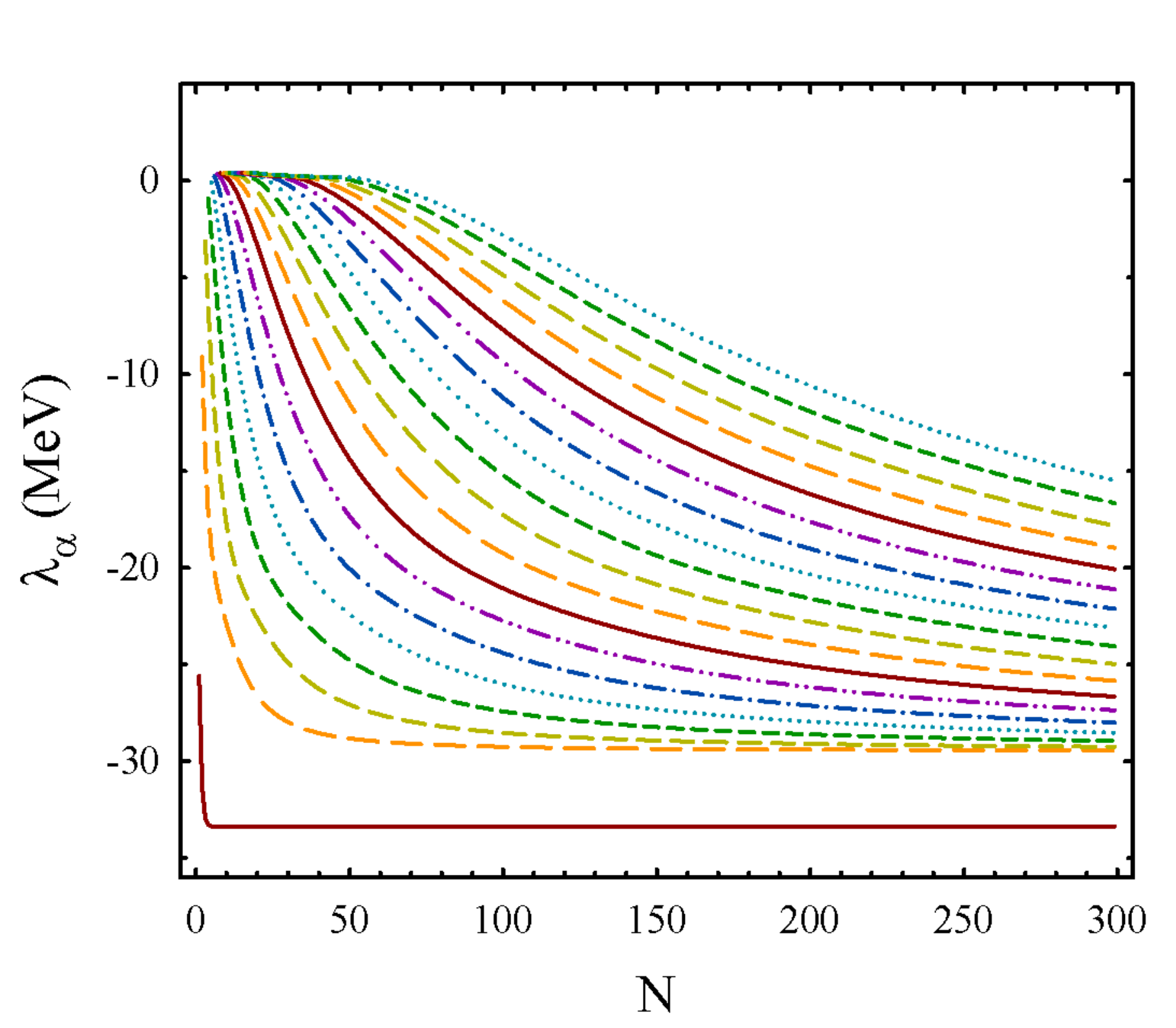}%
\caption{Dependence of eigenvalues of the $0^{+}$ state in $^{6}$Li on the
number of oscillator functions involved in calculations. Results are obtained
with the VP.}%
\label{Fig:EigSpectrvsN6LiL0VP}%
\end{center}
\end{figure}

Fig. \ref{Fig:EigSpectrvsN6LiL0VP}, where the eigenvalues of the $0^+$ state in $^6$Li are shown, 
demonstrates another interesting feature of
the eigenvalues of potential energy operator generated by the Volkov NN potential:
appearance of bound or trapped states. As we can see, the first eigenvalue
$\lambda_{1}$ is almost independent on the number of oscillator functions when
N$\geq5$. For a two-cluster Hamiltonian, such a behavior of eigenvalues
indicates that there is a deeply bound  state with a compact two-cluster
configuration. That is why we called the first eigenvalue $\lambda_{1}$ of the
potential energy matrix a bound or trapped state. Such a behavior of
eigenstates is observed also for the $L^{\pi}=1^{-}$\ states in
$^{5}$He and $^{5}$Li with the MP and VP potentials, for the $L^{\pi}=0^{+}%
$\ states in $^{6}$Li with the MP and VP and for the $L^{\pi}=2^{+}$\ with the
VP. In $^{7}$Li and $^{7}$Be such a state is found for the $1^{-}%
$\ state with the VP only.%

\subsection{Eigenfunctions}

\begin{figure}[ptb]
\begin{center}
\includegraphics[
width=\textwidth] 
{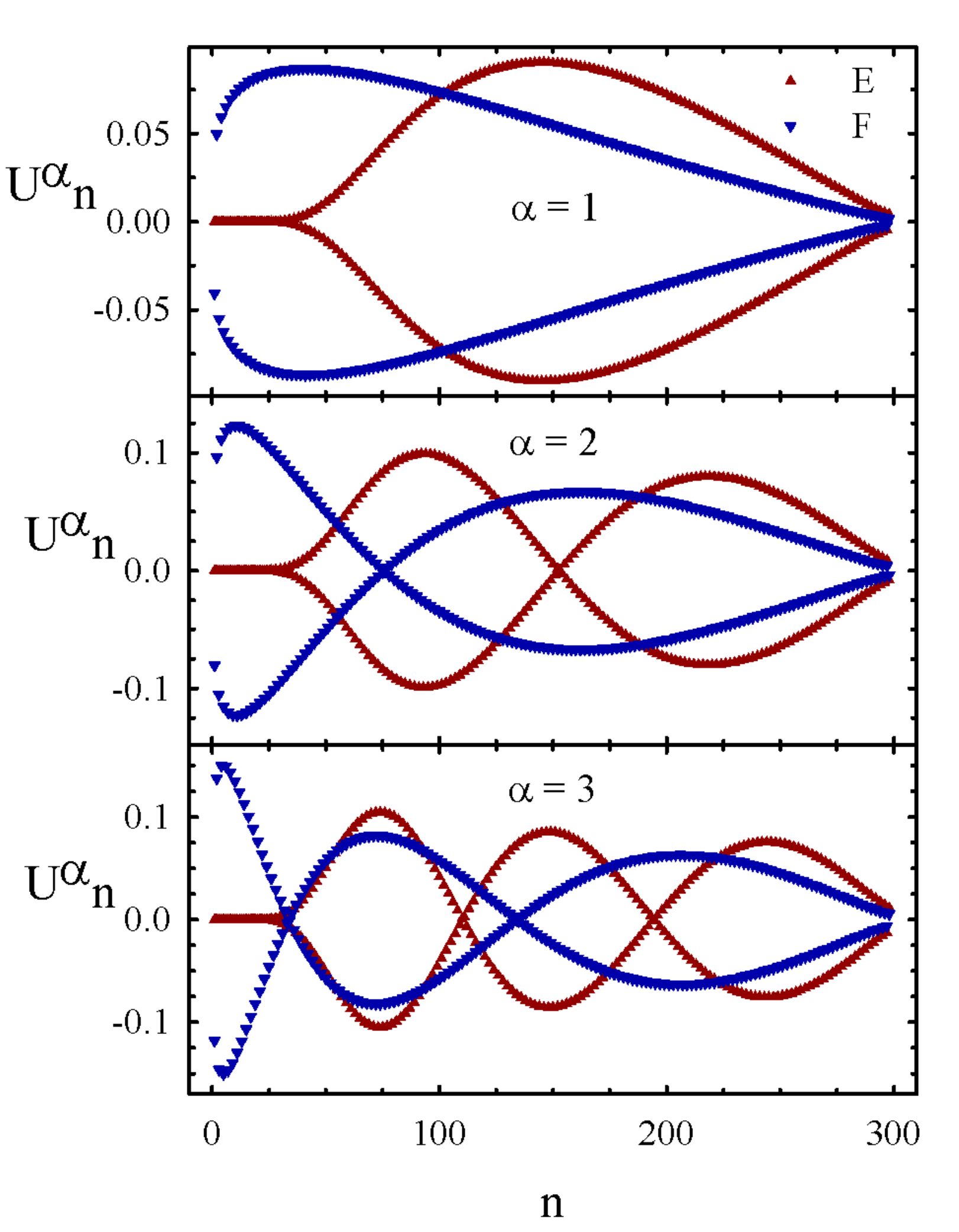}%
\caption{The eigenfunctions of the exact (E) and folding (F) potential energy
in oscillator representation for the $0^{+}$ state in $^{8}$Be. Results are
obtained with the MHNP.}%
\label{Fig:EigFuns8BeL0MHNP}%
\end{center}
\end{figure}

Consider eigenvectors of the potential energy matrix for the $0^{+}$ state in
$^{8}$Be and for the 1$^{-}$ state in $^{7}$Li. In Fig.
\ref{Fig:EigFuns8BeL0MHNP} we compare eigenvectors for $^{8}$Be with and
without antisymmetrization for the MHNP. One can see that they are quite
different. The Pauli principle makes zero the first 50 expansion coefficients
$U_{n}^{\alpha}$.%

\begin{figure}[ptb]
\begin{center}
\includegraphics[
width=\textwidth] 
{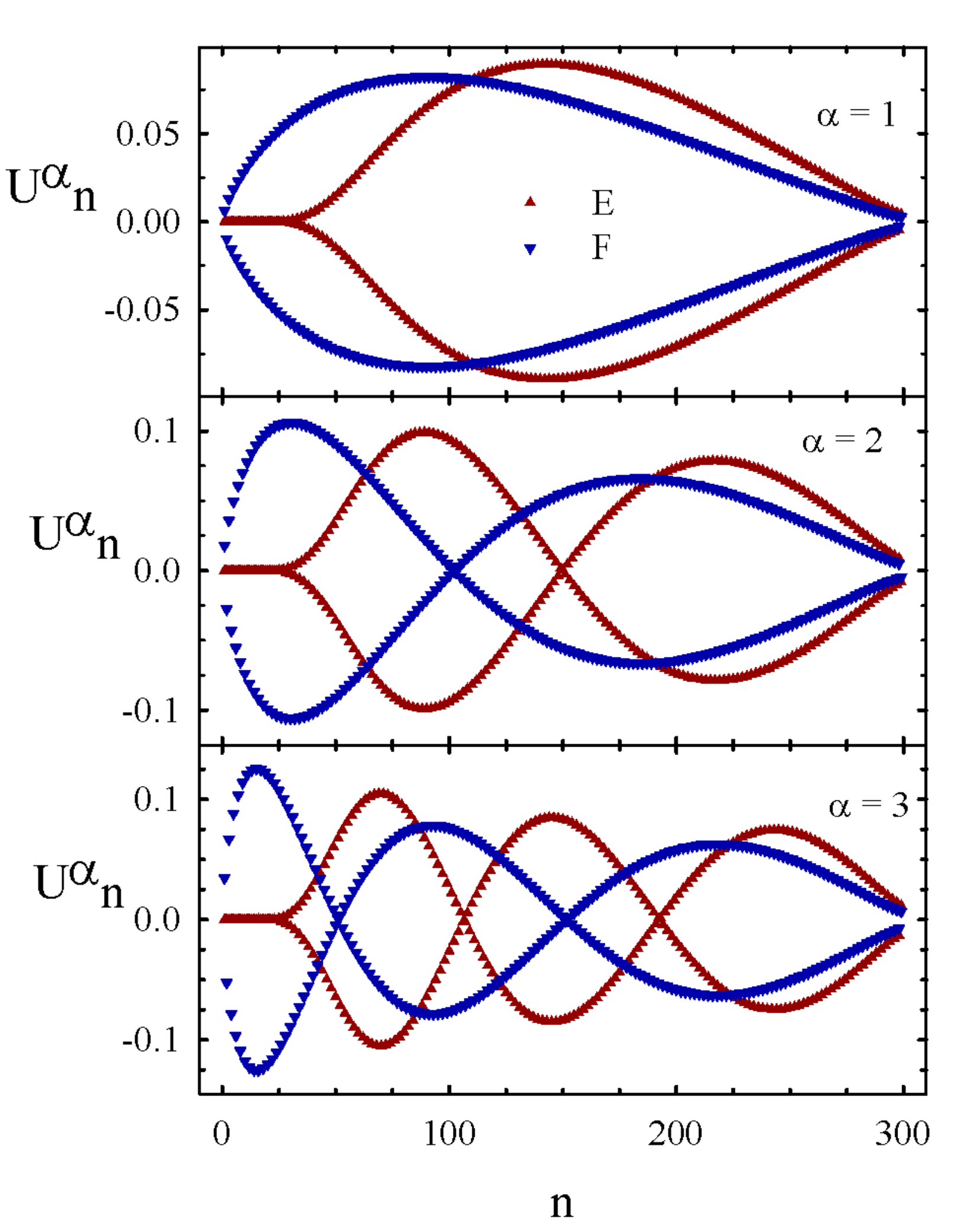}%
\caption{The eigenfunctions of the potential energy matrix generated by the
MHNP for the $1^{-}$ state in $^{7}$Li. }%
\label{Fig:EigFuns7LiL1MHNP}%
\end{center}
\end{figure}

In Fig. \ref{Fig:EigFuns7LiL1MHNP} we observe the same behavior of the
eigenfunctions for the $1^{-}$ state in $^{7}$Li.%

\begin{figure}[ptb]
\begin{center}
\includegraphics[
width=\textwidth] 
{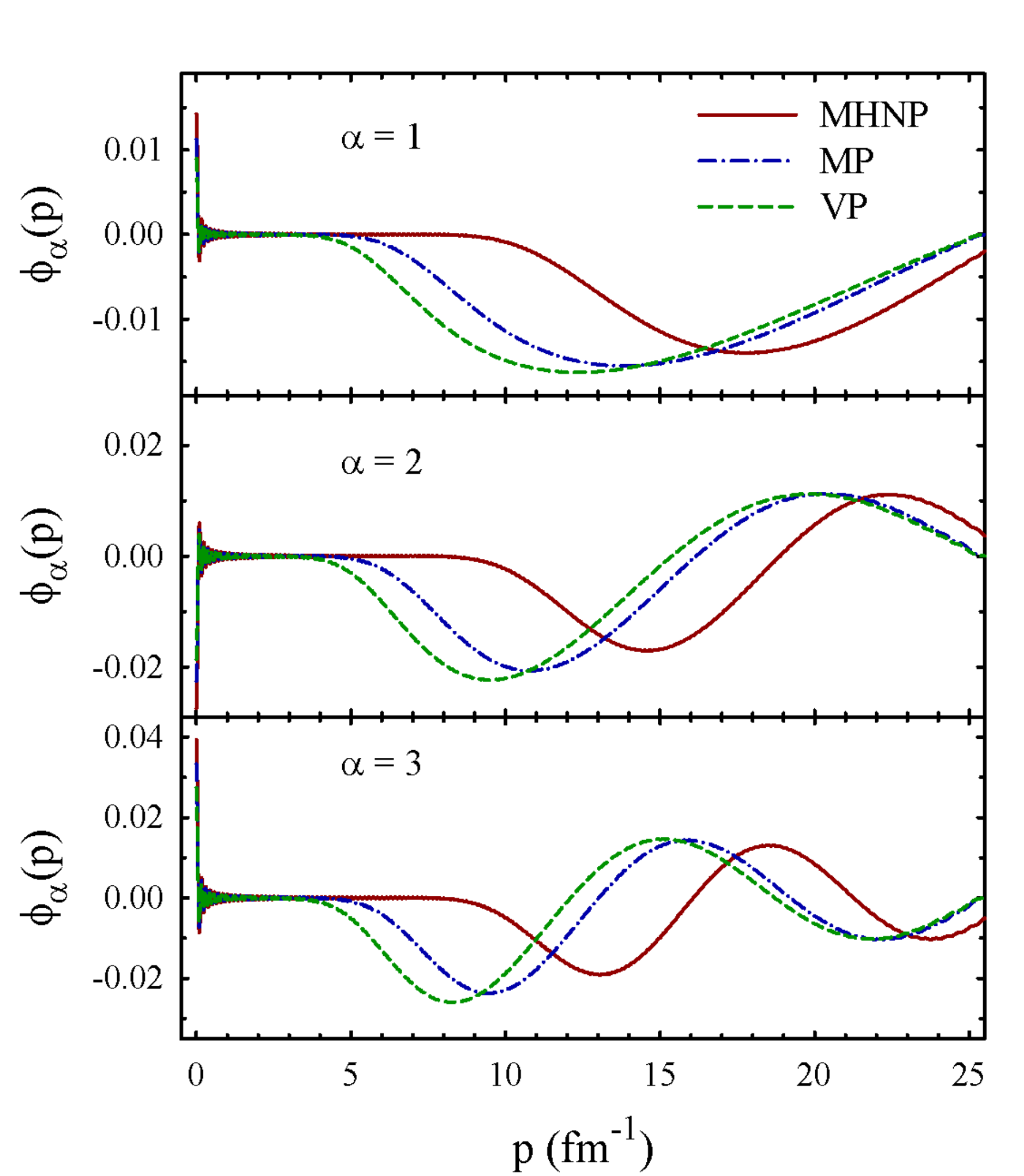}%
\caption{Eigenfunctions of the potential energy matrix obtained for the
$0^{+}$ state in $^{8}$Be with different NN potentials.}%
\label{Fig:EignFuns8BeL0MP}%
\end{center}
\end{figure}

Eigenfunctions of $^{8}$Be obtained with different NN potentials are displayed
in Fig. \ref{Fig:EignFuns8BeL0MP}. We demonstrate the eigenfunctions in the
momentum space for $\alpha$ = 1, 2 and 3. A huge repulsive core in the MHNP
and the Pauli principle make eigenfunctions $\phi_{\alpha}\left(  p\right)  $
\ \textit{to} vanish in a large range of $0<p<10$ fm$^{-1}$, while a rather
modest core in the MP reduces this region to $p\approx$ 6 fm$^{-1}$. The
eigenfunctions $\phi_{\alpha}\left(  p\right)  $ obtained with the VP exhibit
only effects of the Pauli principle, since this potential has a negligibly
small repulsive core.%

\begin{figure}[ptb]
\begin{center}
\includegraphics[
width=\textwidth] 
{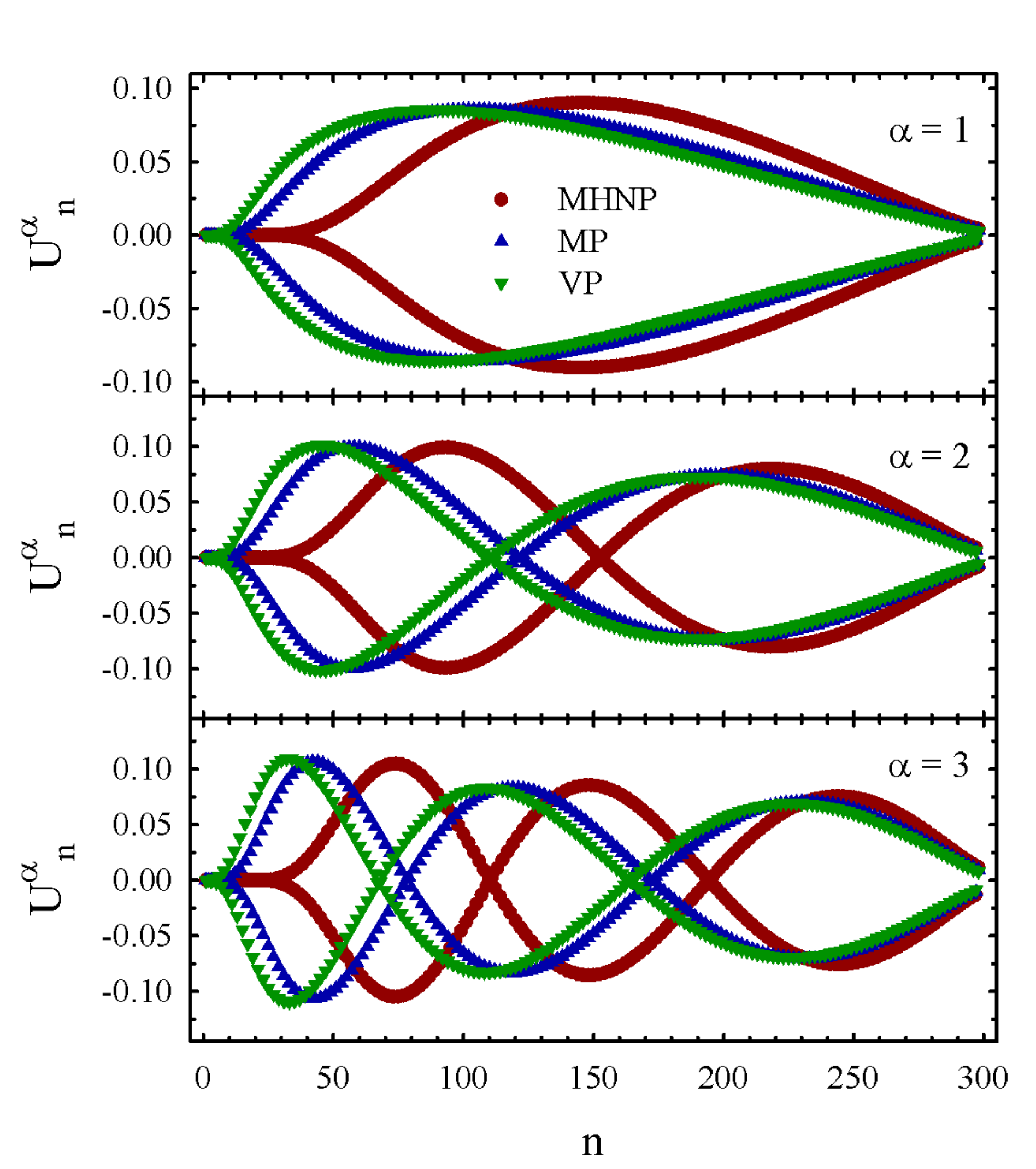}%
\caption{The eigenfunctions of the potential energy operator for the $0^{+}$
state in $^{8}$Be in the oscillator representation.}%
\label{Fig:EignFuns8BeL0CS}%
\end{center}
\end{figure}

Behavior of the eigenfunctions $\phi_{\alpha}\left(  p\right)  $ in the
momentum space is compatible with the behavior of these eigenfunctions\ $U_{n}%
^{\alpha}$ in the oscillator space. Indeed, in Fig. \ref{Fig:EignFuns8BeL0CS}
we observe the same impact of the Pauli principle and a core of the NN
potentials on the eigenfunctions $U_{n}^{\alpha}$.%

By comparing eigenvalues and eigenfunctions of the mirror nuclei $^{5}$He and
$^{5}$Li, $^{7}$Li and $^{7} $Be, we came to the conclusion that the Coulomb
forces have a small impact on the matrix of the potential energy operator and
its eigenvalues and eigenfunctions. 
We do not compare the eigenfunctions of the mirror nuclei as
they are almost undistinguished.

\subsubsection{Resonance and trapped states wave functions.}

\begin{figure}[ptb]
\begin{center}
\includegraphics[
width=\textwidth] 
{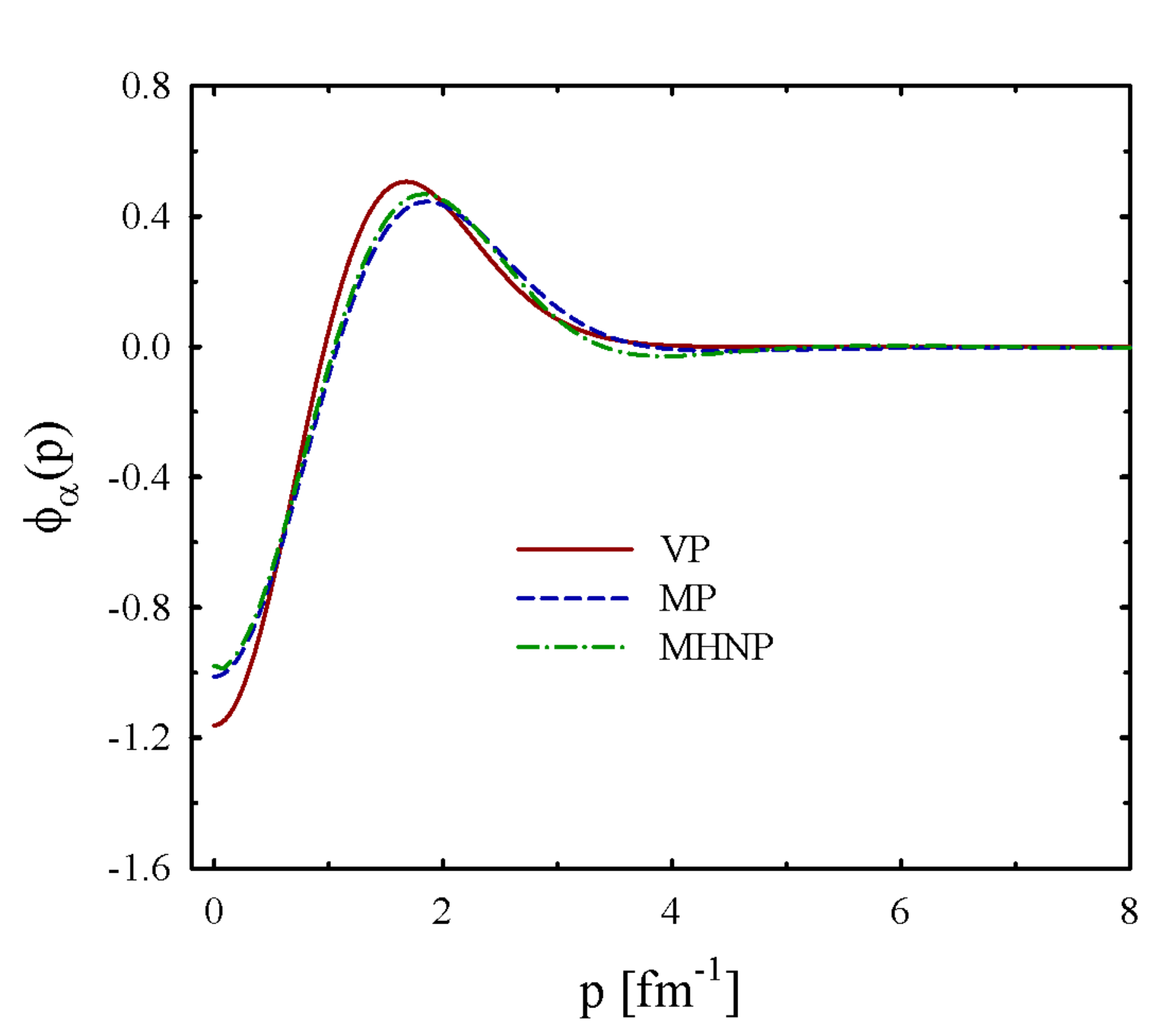}%
\caption{Wave functions of the trapped and resonance $0^{+}$ states in $^{6}%
$Li as a function of momentum $p$.}%
\label{Fig:ResonWF6Li}%
\end{center}
\end{figure}

\begin{figure}[ptb]
\begin{center}
\includegraphics[
width=\textwidth] 
{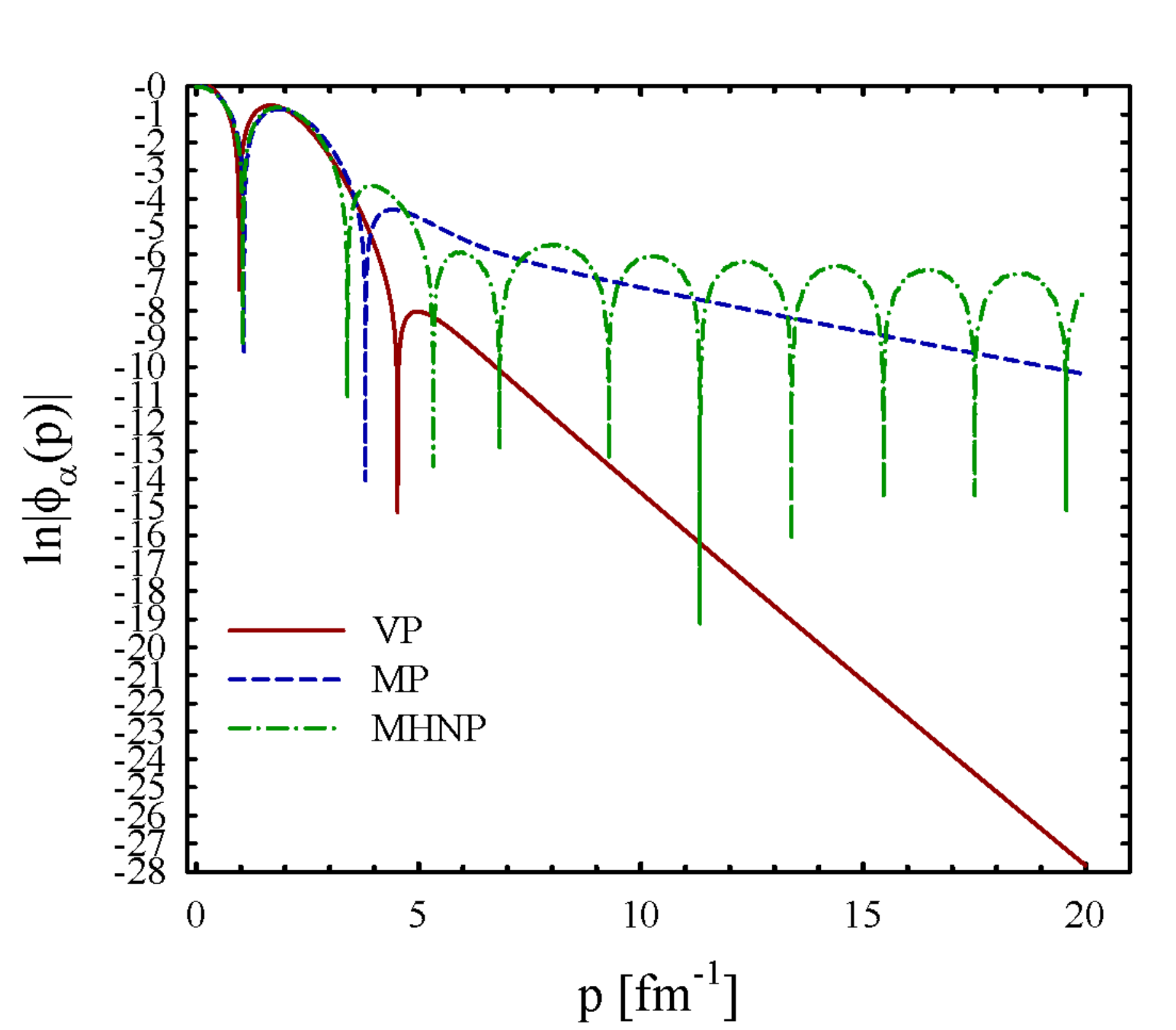}%
\caption{Asymptotic behavior of the wave functions of the $0^{+}$ state in
$^{6}$Li.}%
\label{Fig:ResonWF6LiLS}%
\end{center}
\end{figure}

Now we consider wave functions of bound (trapped) and resonance states in the
two-cluster systems. We start with the wave functions of a trapped $0^{+}$
state in $^{6}$Li. They are displayed in the momentum space in Fig.
\ref{Fig:ResonWF6Li}. One can see that these functions describe a compact
configuration and slightly depend on the shape of the nucleon-nucleon
potentials. Besides, the wave functions obtained with the MP and VP have an
exponential asymptotic behavior (Fig. \ref{Fig:ResonWF6LiLS}). Thus, there is
a full resemblance of these functions with a true bound state wave function
which is usually observed in coordinate space. As for the MHNP, the asymptotic
part of the wave function has an oscillatory behavior (Fig.
\ref{Fig:ResonWF6LiLS}) which is typical for wave functions of a narrow
resonance state. It is important to note that for the VP and MP we have got
the trapped state for this nucleus, they are the lowest eigenstates ($\alpha
$=1) for these potentials. It can be seen from Fig. \ref{Fig:ResonWF6LiLS}
that the VP creates deeper trapped state than the MP, as the wave function
obtained with VP is decreased faster than the wave function generated by the
MP. The resonance state obtained with the MHNP is the $\alpha$=10th
eigenstate. An interesting feature of the wave functions presented in Fig.
\ref{Fig:ResonWF6Li} that they have a node at $p\approx1.0$ fm$^{-1}$. Such a
behavior of eigenfunctions of the trapped state is similar to the behavior of
the wave function in coordinate representation of the ground state of $^{6}%
$Li, obtained within a two-cluster model (see Fig. 5 in Ref.
\cite{2015NuPhA.941..121L}). The node of the bound state wave functions
appears due to the orthogonality of this state to the Pauli forbidden state(s) in
two-cluster systems.%

%

The trapped and resonance states are compact states, because their wave
functions dominates at small values of coordinate ($q$), momentum ($p$) and
oscillator ($n$) spaces.

\begin{figure}[ptb]
\begin{center}
\includegraphics[
width=\textwidth] 
{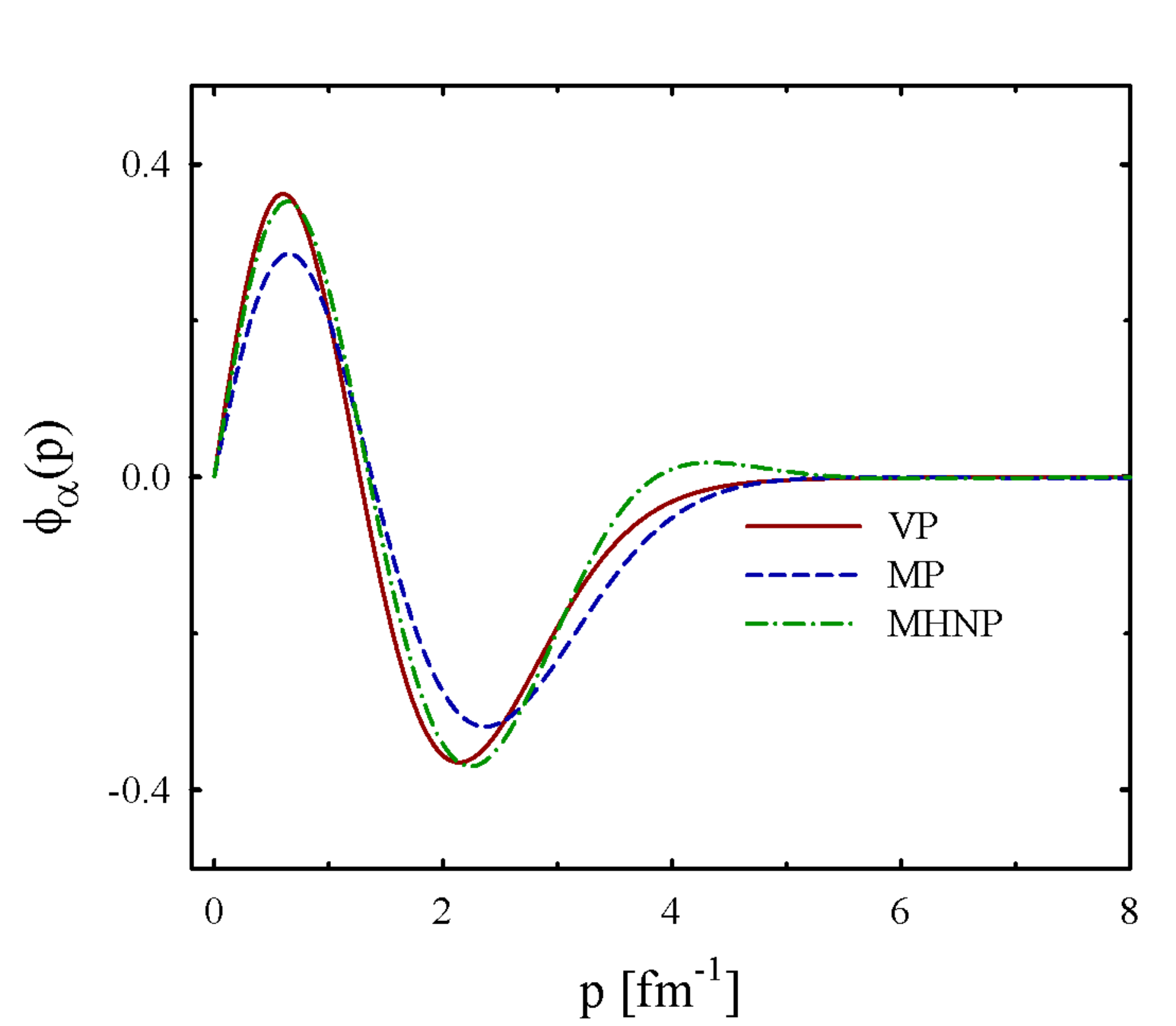}%
\caption{Wave functions of the first $L^{\pi}=1^{-}$ resonance eigenstate in
$^{7}$Li.}%
\label{Fig:ResonFuns7LiMR}%
\end{center}
\end{figure}

Wave functions $\phi_{\alpha}\left(  p\right)  $ of the first $L^{\pi}=1^{-}$
resonance eigenstate in $^{7}$Li generated by different NN potentials are
displayed in Fig. \ref{Fig:ResonFuns7LiMR}. This figure demonstrates that the
shape of these functions slightly depends on the peculiarities of the
nucleon-nucleon potential. The wave functions $\phi_{\alpha}\left(  p\right)
$ equal zero at $p=0$, as these functions describe the relative motion of
clusters with the orbital momentum $L=1$.%

\begin{figure}[ptb]
\begin{center}
\includegraphics[
width=\textwidth] 
{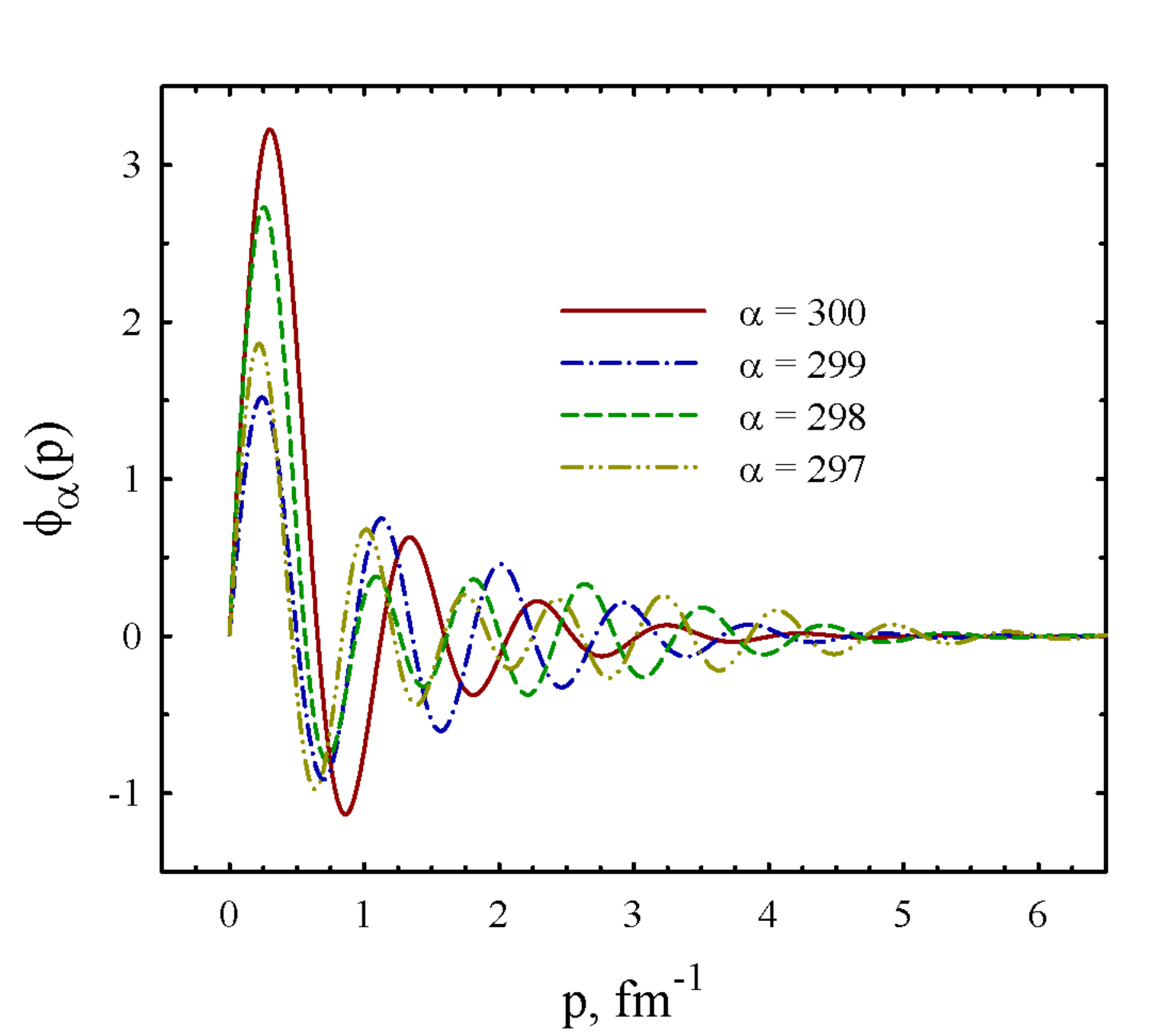}%
\caption{The eigenfunctions $\phi_{\alpha}\left(p\right)$ of the potential
energy matrix corresponding to the highly-excited eigenstates.
These functions are obtained for the $1^{-}$ state in $^{7}$Li with MHNP.}%
\label{Fig:SEigFuns7LiL1MHN}%
\end{center}
\end{figure}

Let us have a closer look at the wave functions of the highly-excited
eigenstates. In Fig. \ref{Fig:SEigFuns7LiL1MHN} we demonstrate the
eigenfunctions $\phi_{\alpha}\left(  p\right)  $ of such states in the
momentum representation. These functions are determined for the\ $1^{-}$ state
in $^{7}$Li with the MHNP. This is the most exotic case with four
highly-excited eigenfunctions. As we see, these functions describe a compact
two-cluster configuration. They are similar to the resonance wave functions.
The compactness of the HES is also observed in the oscillator and coordinate
representations. An asymptotic part of the eigenfunctions $\phi_{\alpha
}\left(  p\right)  $ exhibits an oscillatory behavior with the amplitude which
is much smaller than the amplitude in the internal region. The eigenfunction
for $\alpha=300$ with the largest value of $\lambda_{\alpha}$ corresponds to
the most compact two-cluster configuration. The smaller is the eigenvalue
$\lambda_{\alpha}$, the less compact is the two-cluster system. Or, in other
words, the smaller is the eigenvalue $\lambda_{\alpha}$, the smaller is the
width of such resonance state.%

By comparing Figs. \ref{Fig:ResonFuns7LiMR} and \ref{Fig:SEigFuns7LiL1MHN}, we
see a certain similarity of the low-energy and highly-excited resonance
states. The wave functions of both states have large amplitude at small values
of the momentum $p$. The main difference between these states is that the wave
functions of the highly-excited states have more nodes as they correspond to
higher energy than the low-energy resonance states.

\begin{figure}[ptb]
\begin{center}
\includegraphics[
width=\textwidth] 
{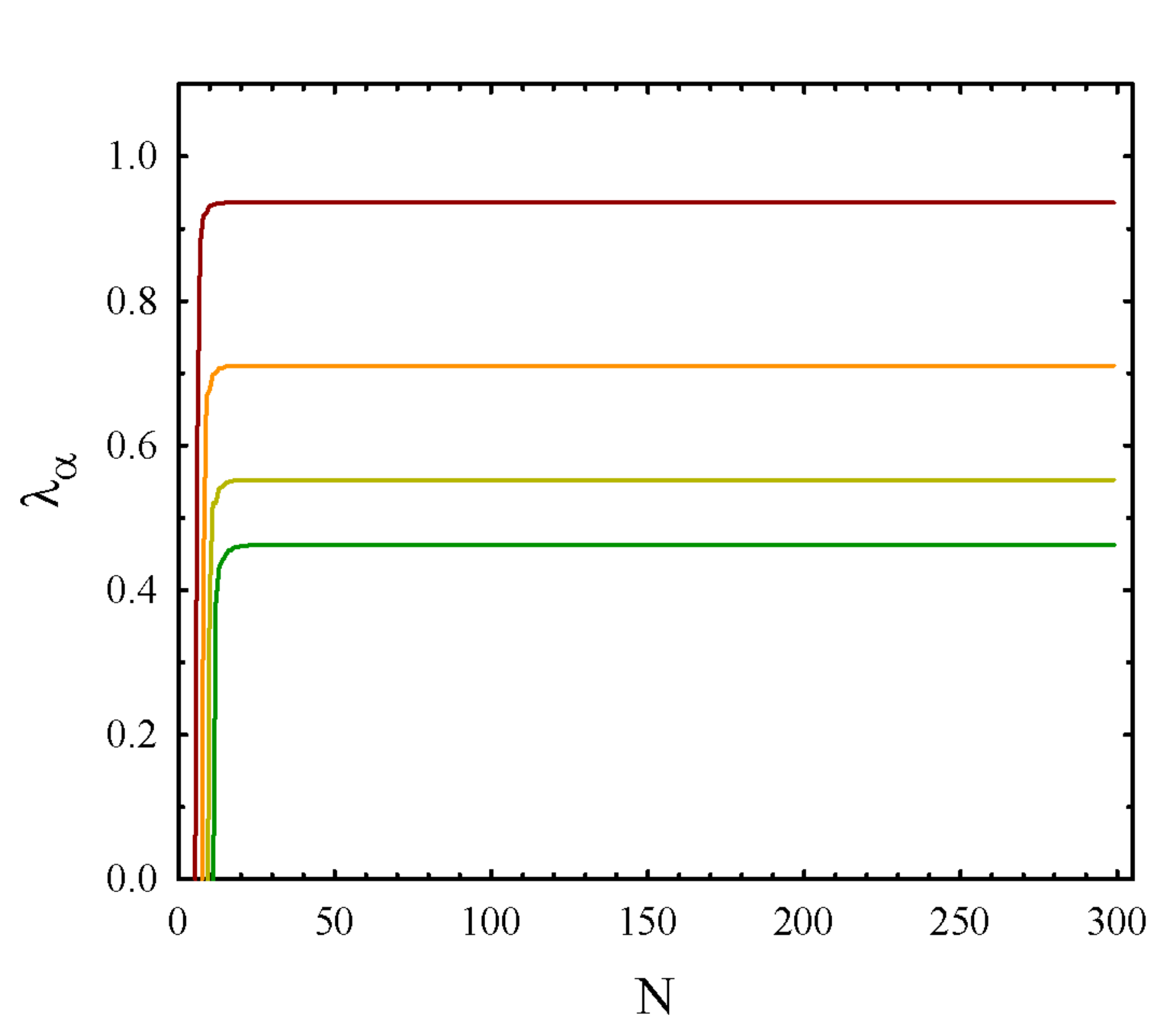}%
\caption{The dependence of the eigenvalues $\lambda_{\alpha}$ belonging to the highly-excited states \ on the
number of oscillator functions involved in calculations. Results are presented
for the $1^{-}$ state in $^{7}$Li and the MHNP.}%
\label{Fig:SEigValvsN7LiL1MHN}%
\end{center}
\end{figure}

If we consider the evolution of the highly excited eigenvalues $\lambda
_{\alpha}$ with the increasing number $N$ of the invoked basis functions, we
observe the stability of the corresponding eigenvalues $\lambda_{\alpha}$,
which is similar to the stability of bound and resonance states displayed in
Figs \ref{Fig:Li7_la_vs_a_L_1_MHNN}, \ref{Fig:EigSpectrvsN6LiL0VP}. As we can
see in Fig. \ref{Fig:SEigValvsN7LiL1MHN}, the last four eigenstates
$\lambda_{\alpha}$ are stable when $N\geq25$.%

\subsection{Interpretation of eigenfunctions}

The shape of the eigenfunctions in the momentum representation suggests that
they are wave functions of a particle in a field of a step potential. The
range of such a potential should be approximately of 7-10 fm. And the height
of the step potential should be large enough to suppress the wave function
inside this potential. However, appearance of "resonance states" or "trapped
states" somewhat contradicts this suggestion. It is known, see, for example,
book \cite{kn:Fluegge71}, that the step potential creates resonance states
with the energy which is larger than the height of the potential. The step
potential cannot create a very narrow resonance states. Besides, the nodes of
wave functions of the trapped or resonance states has to be taken into account.
To establish the explicit form of the auxiliary Hamiltonian $\widehat{H}_{a}$
\begin{equation}
\widehat{H}_{a}=\widehat{T}_{p}+\widehat{V}_{a}\left(  p\right)  ,
\label{eq:400}%
\end{equation}
which possess the following property%
\[
\widehat{H}_{a}\phi_{\alpha}\left(  p\right)  =\mathcal{E}_{\alpha}%
\phi_{\alpha}\left(  p\right)  ,
\]
one needs to perform a large number of numerical experiments. Thus this
problem will be considered elsewhere.

\section{Conclusions \label{Sec:Conclusions}}

We have investigated effects of antisymmetrization on the potential energy of
two clusters interaction. For this aim we have studied eigenfunctions and
eigenvalues of the potential energy matrix, calculated within a basis of the
cluster oscillator functions. It was shown that eigenvalues of the potential
energy calculated with the full antisymmetrization are close to the ones,
obtained in the folding approximation. However, the eigenfunctions in those
cases are quite different. In the folding model the eigenfunctions are the
free-motion Bessel functions, while the antisymmetrization generates functions
which are similar to the functions, describing scattering states on a step
potential. However, to establish the exact nature of \ these eigenfunctions
one needs to perform enhanced investigations. This is planned to do in the
next publication.

By analyzing the evolution of the eigenvalues of the potential energy operator
with increasing the number $N$ of oscillator functions, we have discovered the
trapped and resonance states. The trapped states represent themselves as a
state with a stable and almost independent on $N$ the lowest eigenvalue
$\lambda_{1}$. The resonance states exhibit themselves as a plateau in the
dependence of the eigenvalues $\lambda_{\alpha}$ on the number $N$of the basis
functions involved in the calculations. Both trapped and resonance states have
a compact wave function of two-cluster systems. This compactness is observed
in the oscillator, momentum and coordinate spaces. The trapped states have an
exponential asymptotic tail, while the resonance states have an oscillating
asymptotic tail.

We have also discovered the highly-excited resonance states. They have the
largest eigenvalues and mainly exhibit themselves in the normal parity states
in all nuclei but $^{8}$Be. The largest number (four) of the highly-excited
resonance states are observed in $^{7}$Li and $^{7}$Be and only one state is
found in other nuclei. The highly-excited states describe a compact
two-cluster structure which can be seen in the oscillator, coordinate and
momentum representations.

By closing the present paper we would like to indicate the further steps of
development of the suggested method and obtained results.

First, the obtained eigenfunctions and eigenvalues of the matrix of potential
energy operator can be used to construct the wave functions and t-matrix for
two-cluster systems in a similar way as it was done in Ref.
\cite{LASHKO2019167930} for two-body potentials. This may suggests an
alternative form of the dynamic equations to that which is now used within the
algebraic version of the resonating group method. This is definitely a subject
for a separate paper.

Second, it is also interesting to study properties of the potential energy
operator for three-cluster system when a several two-cluster channels and a
large number of three-cluster channels are open. This will be studied in other papers.

\section{Acknowledgment}

The present work was partially supported by the Program of Fundamental
Research of the Department of Physics and Astronomy of the National Academy of
Sciences of Ukraine (project No. 0117U000239).


\begin{thebibliography}{10}
\expandafter\ifx\csname url\endcsname\relax
  \def\url#1{\texttt{#1}}\fi
\expandafter\ifx\csname urlprefix\endcsname\relax\def\urlprefix{URL }\fi
\expandafter\ifx\csname href\endcsname\relax
  \def\href#1#2{#2} \def\path#1{#1}\fi

\bibitem{kn:Fil_Okhr}
G.~F. Filippov, I.~P. Okhrimenko, Use of an oscillator basis for solving
  continuum problems, Sov. J. Nucl. Phys. {\bf 32} (1981) 480--484.

\bibitem{kn:Fil81}
G.~F. Filippov, On taking into account correct asymptotic behavior in
  oscillator-basis expansions, Sov. J. Nucl. Phys. {\bf 33} (1981) 488--489.

\bibitem{1937PhRv...52.1083W}
J.~A. {Wheeler}, {Molecular Viewpoints in Nuclear Structure}, Phys. Rev. 52
  (1937) 1083--1106.
\newblock \href {http://dx.doi.org/10.1103/PhysRev.52.1083}
  {\path{doi:10.1103/PhysRev.52.1083}}.

\bibitem{1937PhRv...52.1107W}
J.~A. {Wheeler}, {On the Mathematical Description of Light Nuclei by the Method
  of Resonating Group Structure}, Phys. Rev. 52 (1937) 1107--1122.
\newblock \href {http://dx.doi.org/10.1103/PhysRev.52.1107}
  {\path{doi:10.1103/PhysRev.52.1107}}.

\bibitem{kn:Saito77}
S.~Saito, {Theory of Resonating Group Method and Generator Coordinate Method,
  and Orthogonality Condition Model}, Prog. Theor. Phys. Suppl. {\bf 62} (1977)
  11--89.
\newblock \href {http://dx.doi.org/10.1143/PTPS.62.11}
  {\path{doi:10.1143/PTPS.62.11}}.

\bibitem{1978PhR....47..167T}
Y.~C. {Tang}, M.~{Lemere}, D.~R. {Thompsom}, {Resonating-group method for
  nuclear many-body problems}, Phys. Rep. 47 (1978) 167--223.
\newblock \href {http://dx.doi.org/10.1016/0370-1573(78)90175-8}
  {\path{doi:10.1016/0370-1573(78)90175-8}}.

\bibitem{1981LNP...145..571T}
Y.~C. {Tang}, {Microscopic description of the nuclear cluster theory}, in:
  T.~S. {Kuo}, S.~M. {Wong} (Eds.), Topics in Nuclear Physics II A
  Comprehensive Review of Recent Developments, Vol. 145 of Lecture Notes in
  Physics, Berlin Springer Verlag, 1981, pp. 571--692.
\newblock \href {http://dx.doi.org/10.1007/BFb0017230}
  {\path{doi:10.1007/BFb0017230}}.

\bibitem{kn:wilderm_eng}
K.~Wildermuth, Y.~Tang, A unified theory of the nucleus, Vieweg Verlag,
  Braunschweig, 1977.

\bibitem{1970NuPhA.145..593K}
I.~V. {Kurdyumov}, Y.~F. {Smirnov}, K.~V. {Shitikova}, S.~K.~E. {Samarai},
  {Translationally invariant shell model}, Nucl. Phys. A 145 (1970) 593--612.
\newblock \href {http://dx.doi.org/10.1016/0375-9474(70)90444-6}
  {\path{doi:10.1016/0375-9474(70)90444-6}}.

\bibitem{Neudachin69E}
{V.~G. Neudachin, Yu.~F. Smirnov}, Nucleon associations in light nuclei (in
  Russian), 'Nauka', Moscow, 1969.

\bibitem{RevaiPrepr1975}
J.~Revai, A new method of calculating wave functions on harmonic ocillator
  basis, Preprint JINR, Dubna, E4-9429 (1975) 12 pp.

\bibitem{kn:majl}
L.~Majling, J.~Rizek, Z.~Pluhar, Y.~F. Smirnov, On some perculiarities of the
  variational calculations in the harmonic oscillator basis, J.~Phys. G:
  Nucl.~Phys. {\bf 2} (1976) 357--364.

\bibitem{kn:moshin}
M.~Moshinsky, The Harmonic oscillator in Modern Physics: From Atoms to Quarks,
  Gordon Breach, New-York, London, Paris, 1969.

\bibitem{1977NuPhA.291..230B}
D.~{Baye}, P.-H. {Heenen}, M.~{Libert-Heinemann}, {Microscopic R-matrix theory
  in a generator coordinate basis (III). Multi-channel scattering}, Nucl. Phys.
  A 291 (1977) 230--240.
\newblock \href {http://dx.doi.org/10.1016/0375-9474(77)90208-1}
  {\path{doi:10.1016/0375-9474(77)90208-1}}.

\bibitem{2009PhRvC..79d4606Q}
S.~{Quaglioni}, P.~{Navr{\'a}til}, {Ab initio many-body calculations of
  nucleon-nucleus scattering}, Phys. Rev. C 79~(4) (2009) 044606.
\newblock \href {http://arxiv.org/abs/0901.0950} {\path{arXiv:0901.0950}},
  \href {http://dx.doi.org/10.1103/PhysRevC.79.044606}
  {\path{doi:10.1103/PhysRevC.79.044606}}.

\bibitem{2013PhRvC..88c4320Q}
S.~{Quaglioni}, C.~{Romero-Redondo}, P.~{Navr{\'a}til}, {Three-cluster dynamics
  within an ab initio framework}, Phys. Rev. C 88~(3) (2013) 034320.
\newblock \href {http://arxiv.org/abs/1307.8160} {\path{arXiv:1307.8160}},
  \href {http://dx.doi.org/10.1103/PhysRevC.88.034320}
  {\path{doi:10.1103/PhysRevC.88.034320}}.

\bibitem{Saito69}
S.~Saito, {Interaction between Clusters and Pauli Principle}, Prog. Theor.
  Phys. {\bf 41}~(3) (1969) 705--722.
\newblock \href {http://dx.doi.org/10.1143/PTP.41.705}
  {\path{doi:10.1143/PTP.41.705}}.

\bibitem{Kukulin:1976vf}
V.~I. Kukulin, V.~G. Neudachin, V.~N. Pomerantsev, {Exclusion of Occupied
  States in the Faddeev Equations for Three Composite Particles}, Yad. Fiz. 24
  (1976) 298--307.

\bibitem{1976TMP....27..549K}
V.~I. {Kukulin}, V.~N. {Pomerantsev}, {Rearrangement and improvement of
  convergence of the born series in scattering theory on the basis of
  orthogonal projections}, Theor. Math. Phys. 27 (1976) 549--557.
\newblock \href {http://dx.doi.org/10.1007/BF01028623}
  {\path{doi:10.1007/BF01028623}}.

\bibitem{1978AnPhy.111..330K}
V.~I. {Kukulin}, V.~N. {Pomerantsev}, {The orthogonal projection method in
  scattering theory}, Ann. Phys. 111 (1978) 330--363.
\newblock \href {http://dx.doi.org/10.1016/0003-4916(78)90069-6}
  {\path{doi:10.1016/0003-4916(78)90069-6}}.

\bibitem{1980PThPh..63..895F}
Y.~{Fujiwara}, H.~{Horiuchi}, {Generator Coordinate Theory of Normalization
  Kernels of Cluster Systems. I}, Prog. Theor. Phys. 63 (1980) 895--918.

\bibitem{1977PThPh..58..204H}
H.~{Horiuchi}, {Multi-Cluster Allowed States and Spectroscopic Amplitude of
  Cluster Transfer}, Prog. Theor. Phys. 58 (1977) 204--222.

\bibitem{1981PThPh..65.1632F}
Y.~{Fujiwara}, H.~{Horiuchi}, {Generator Coordinate Theory of Normalization
  Kernels of Cluster Systems. II ---Systems Described by Three Generator
  Coordinate Vectors Involving Complex Conjugate One(s)---}, Prog. Theor. Phys.
  65 (1981) 1632--1666.

\bibitem{1981PThPh..65.1901F}
Y.~{Fujiwara}, H.~{Horiuchi}, {Generator Coordinate Theory of Normalization
  Kernels of Cluster Systems. III ---Application of Double Gel'fand Polynomials
  to General Cluster Systems---}, Prog. Theor. Phys. 65 (1981) 1901--1927.

\bibitem{1983PThPh..70..809F}
Y.~{Fujiwara}, Y.~C. {Tang}, H.~{Horiuchi}, {Generator Coordinate Theory of
  Normalization Kernels of Cluster Systems. IV ---Application of Double
  Gel'fand Polynomials to $SU_{4}$ Symmetry of Cluster Wave Functions---},
  Prog. Theor. Phys. 70 (1983) 809--826.
\newblock \href {http://dx.doi.org/10.1143/PTP.70.809}
  {\path{doi:10.1143/PTP.70.809}}.

\bibitem{1984PThPh..72.1277H}
H.~{Horiuchi}, K.~{Yabana}, {Pauli-Forbidden Region in the Phase-Space of
  Coupled-Channel System in the Framework of the Time-Dependent Variational
  Theory}, Prog. Theor. Phys. 72 (1984) 1277--1281.

\bibitem{1988PThPh..80..663K}
K.~{Kato}, K.~{Fukatsu}, H.~{Tanaka}, {Systematic Construction Method of
  Multi-Cluster Pauli-Allowed States}, Prog. Theor. Phys. 80~(4) (1988)
  663--677.
\newblock \href {http://dx.doi.org/10.1143/PTP.80.663}
  {\path{doi:10.1143/PTP.80.663}}.

\bibitem{2003FBS....33..173F}
G.~F. {Filippov}, Y.~A. {Lashko}, S.~V. {Korennov}, K.~{Kat{\= o}}, {Norm
  Kernels and the Closeness Relation for Pauli-Allowed Basis Functions},
  Few-Body Syst. 33 (2003) 173--198.
\newblock \href {http://dx.doi.org/10.1007/s00601-003-0009-z}
  {\path{doi:10.1007/s00601-003-0009-z}}.

\bibitem{2004PhRvC..70f4001F}
G.~{Filippov}, Y.~{Lashko}, {Peculiar properties of the cluster-cluster
  interaction induced by the Pauli exclusion principle}, Phys. Rev. C 70~(6)
  (2004) 064001.
\newblock \href {http://dx.doi.org/10.1103/PhysRevC.70.064001}
  {\path{doi:10.1103/PhysRevC.70.064001}}.

\bibitem{2005EChAYa..36.1373F}
G.~{Filippov}, Y.~{Lashko}, {Structure of Light Neutron-Rich Nuclei and Nuclear
  Reactions Involving These Nuclei}, El. Chast. Atom. Yadra 36~(6) (2005)
  1373--1424.

\bibitem{2009NuPhA.826...24L}
Y.~A. {Lashko}, G.~F. {Filippov}, {The role of the Pauli principle in
  three-cluster systems composed of identical clusters}, Nucl. Phys. A 826
  (2009) 24--48.
\newblock \href {http://arxiv.org/abs/0811.1695} {\path{arXiv:0811.1695}},
  \href {http://dx.doi.org/10.1016/j.nuclphysa.2009.05.071}
  {\path{doi:10.1016/j.nuclphysa.2009.05.071}}.

\bibitem{2008NuPhA.806..124L}
Y.~A. {Lashko}, G.~F. {Filippov}, {How the Pauli principle governs the decay of
  three-cluster systems}, Nucl. Phys. A 806 (2008) 124--145.
\newblock \href {http://arxiv.org/abs/0712.4013} {\path{arXiv:0712.4013}},
  \href {http://dx.doi.org/10.1016/j.nuclphysa.2008.03.003}
  {\path{doi:10.1016/j.nuclphysa.2008.03.003}}.

\bibitem{LASHKO2019167930}
Y.~A. Lashko, V.~Vasilevsky, G.~Filippov,
  \href{http://www.sciencedirect.com/science/article/pii/S000349161930185X}{Properties
  of a potential energy matrix in oscillator basis}, Ann. Phys. 409 (2019)
  167930.
\newblock \href {http://dx.doi.org/https://doi.org/10.1016/j.aop.2019.167930}
  {\path{doi:https://doi.org/10.1016/j.aop.2019.167930}}.

\bibitem{2005PPN..36.714F}
G.~{Filippov}, Y.~{Lashko}, {Structure of Light Neutron-Rich Nuclei and Nuclear
  Reactions Involving These Nuclei}, Phys. Part. Nucl. 36~(6) (2005) 714--739.

\bibitem{kn:cohstate2E}
G.~F. Filippov, V.~S. Vasilevsky, L.~L. Chopovsky, Solution of problems in the
  microscopic theory of the nucleus using the technique of generalized coherent
  states, Sov. J. Part. Nucl. {\bf 16} (1985) 153--177.

\bibitem{kn:cohstate1E}
G.~F. Filippov, V.~S. Vasilevsky, L.~L. Chopovsky, Generalized coherent states
  in nuclear-physics problems, Sov. J. Part. Nucl. {\bf 15} (1984) 600--619.

\bibitem{1968PThPh..40..893S}
S.~{Saito}, {Effect of Pauli Principle in Scattering of Two Clusters}, Prog.
  Theor. Phys. 40 (1968) 893--894.
\newblock \href {http://dx.doi.org/10.1143/PTP.40.893}
  {\path{doi:10.1143/PTP.40.893}}.

\bibitem{Zubarev_EChAYa76}
A.~Zubarev, Separabilization method in the problems of nuclear physics,
  El.~Chast.~Atom.~Jadra {\bf 7}~(2) (1976) 553--583.

\bibitem{bookBelyaev786E}
V.~B. Belyaev, Lectures on theory of few body systems. (In Russian),
  Energoatomizdat, Moscow, 1986.

\bibitem{kn:Newton}
R.~G. Newton, Scattering Theory of Waves and Particles, McGraw-Hill, New-York,
  1966.

\bibitem{kn:Volk65}
A.~B. Volkov, Equilibrum deformation calculation of the ground state energies
  of 1p shell nuclei, Nucl. Phys. {\bf 74} (1965) 33--58.
\newblock \href {http://dx.doi.org/10.1016/0029-5582(65)90244-0}
  {\path{doi:10.1016/0029-5582(65)90244-0}}.

\bibitem{potMHN1}
A.~Hasegawa, S.~Nagata, Ground state of \makebox{$^6$Li}, Prog. Theor. Phys.
  {\bf 45} (1971) 1786--1807.
\newblock \href {http://dx.doi.org/10.1143/PTP.45.1786}
  {\path{doi:10.1143/PTP.45.1786}}.

\bibitem{potMHN2}
F.~Tanabe, A.~Tohsaki, R.~Tamagaki, $\alpha \alpha$ scattering at intermediate
  energies, Prog. Theor. Phys. {\bf 53} (1975) 677--691.

\bibitem{kn:Minn_pot1}
D.~R. Thompson, M.~LeMere, Y.~C. Tang, Systematic investigation of scattering
  problems with the resonating-group method, Nucl. Phys. {\bf A286}~(1) (1977)
  53--66.
\newblock \href {http://dx.doi.org/10.1016/0375-9474(77)90007-0}
  {\path{doi:10.1016/0375-9474(77)90007-0}}.

\bibitem{1967PhRvL..19..173H}
F.~E. {Harris}, {Expansion Approach to Scattering}, Phys. Rev. Lett. 19~(4)
  (1967) 173--175.
\newblock \href {http://dx.doi.org/10.1103/PhysRevLett.19.173}
  {\path{doi:10.1103/PhysRevLett.19.173}}.

\bibitem{1984JMP....25..317Y}
H.~A. {Yamani}, {The J-matrix reproducing kernel: Numerical weights at the
  Harris energy eigenvalues}, J. Math. Phys. 25 (1984) 317--322.
\newblock \href {http://dx.doi.org/10.1063/1.526152}
  {\path{doi:10.1063/1.526152}}.

\bibitem{1970PhRvA...1.1109H}
A.~U. {Hazi}, H.~S. {Taylor}, {Stabilization Method of Calculating Resonance
  Energies: Model Problem}, Phys. Rev. A 1 (1970) 1109--1120.
\newblock \href {http://dx.doi.org/10.1103/PhysRevA.1.1109}
  {\path{doi:10.1103/PhysRevA.1.1109}}.

\bibitem{2015NuPhA.941..121L}
Y.~A. {Lashko}, G.~F. {Filippov}, V.~S. {Vasilevsky}, {Dynamics of two-cluster
  systems in phase space}, Nucl. Phys. A 941 (2015) 121--144.
\newblock \href {http://arxiv.org/abs/1503.06005} {\path{arXiv:1503.06005}},
  \href {http://dx.doi.org/10.1016/j.nuclphysa.2015.06.006}
  {\path{doi:10.1016/j.nuclphysa.2015.06.006}}.

\bibitem{kn:Fluegge71}
S.~Fluegge, Practical quantum mechanics, Springer, Berlin, 1971.

\end{thebibliography}
\end{document}